\documentclass[english,aps,nofootinbib,longbibliography,notitlepage,superscriptaddress,prd,preprintnumbers,showkeys,twocolumn]{revtex4-1}

\usepackage[usenames,dvipsnames,svgnames,table]{xcolor}
\usepackage{verbatim}
\usepackage[T1]{fontenc}
\usepackage[latin9]{inputenc}
\usepackage{babel}
\usepackage{float}
\usepackage{graphicx}
\usepackage{svg}
\usepackage{hyperref}
\usepackage{amsthm,amsmath,amssymb,amsfonts}
\usepackage[mathscr]{eucal}
\usepackage{microtype}
\usepackage{esint}
\usepackage{amsbsy}
\usepackage{color}
\usepackage[capitalise]{cleveref}
\usepackage{breakurl}
\usepackage[normalem]{ulem}
\usepackage{bbm}
\usepackage{mathtools}
\usepackage{braket}
\usepackage{comment}
\usepackage[super]{nth}
\usepackage{csquotes}
\usepackage{lmodern}
\usepackage{hyphenat}

\definecolor{lightgray}{gray}{0.9}
\definecolor{Amber}{rgb}{1.0, 0.75, 0.0}
\definecolor{blizzardblue}{rgb}{0.67, 0.9, 0.93}
\definecolor{myblue}{HTML}{0273B2}
\colorlet{myblueLight}{myblue!35}

\usepackage{tikz}
\usepackage{pgfplots}
\pgfplotsset{compat=newest,every axis plot/.append style={line width=1pt}}

\allowdisplaybreaks

\makeatletter

\pdfpageheight\paperheight
\pdfpagewidth\paperwidth

\@ifundefined{textcolor}{}{%
 \definecolor{BLACK}{gray}{0}
 \definecolor{WHITE}{gray}{1}
 \definecolor{RED}{rgb}{1,0,0}
 \definecolor{GREEN}{rgb}{0,1,0}
 \definecolor{BLUE}{rgb}{0,0,1}
 \definecolor{CYAN}{cmyk}{1,0,0,0}
 \definecolor{MAGENTA}{cmyk}{0,1,0,0}
 \definecolor{YELLOW}{cmyk}{0,0,1,0}
}

\makeatother

\makeatletter

\DeclareRobustCommand{\rcite}[1]{%
  \rcite@aux#1,\@nil{#1}%
}
\def\rcite@aux#1,#2\@nil#3{%
  \if\relax#2\relax
    Ref.~\cite{#3}%
  \else
    Refs.~\cite{#3}%
  \fi
}
\makeatother
\definecolor{wine}{RGB}{136,34,85}
\definecolor{teal}{RGB}{0,85,102}
\hypersetup{
    colorlinks = true,
    citecolor = {wine},
    linkcolor = {teal},
    urlcolor = {teal},
}

\newcommand{\be}{\begin{equation}}
\newcommand{\ee}{\end{equation}}
\newcommand{\ba}{\begin{eqnarray}}
\newcommand{\ea}{\end{eqnarray}}

\def\MPl{M_{\rm{Pl}}}

\def\half{\frac{1}{2}}

\def\rhom{\rho_{\rm{m}}}

\def\rhoDE{\rho_{\rm{DE}}}
\def\PDE{P_{\rm{DE}}}
\def\cH{\mathcal{H}}
\def\Ta{\mathcal{T}}
\def\N{\mathcal{N}}
\def\G{\mathcal{G}}
\def\csm2{c_{\rm{sl}}^2}
\def\css2{c_{\rm{ss}}^2}
\def\csr2{c_{\rm{sr}}^2}

\begin{document}

\preprint{IFT-UAM/CSIC-26-85}

\title{Cosmological perturbations and clustering mechanisms in multifield dark energy}

\author{Mohammad Arab}
\email{m.arab92@basu.ac.ir}
\affiliation{Department of Physics, Faculty of Science, Bu-Ali Sina University, Hamedan 65178, Iran}

\author{Yashar Akrami}
\email{yashar.akrami@csic.es}
\affiliation{Instituto de F\'isica Te\'orica (IFT) UAM-CSIC, C/ Nicol\'as Cabrera 13-15, Campus de Cantoblanco UAM, 28049 Madrid, Spain}
\affiliation{CERCA/ISO, Department of Physics, Case Western Reserve University, 10900 Euclid Avenue, Cleveland, OH 44106, USA}
\affiliation{Astrophysics Group \& Imperial Center for Inference and Cosmology, Department of Physics, Imperial College London, Blackett Laboratory, Prince Consort Road, London SW7 2AZ, United Kingdom}

\date{\today}

\begin{abstract}
We investigate the cosmological signatures of two-field dark energy with curved field-space geometry. We numerically solve the background and linear perturbation equations and assess the validity of the effective single-field description. We examine three distinct dark-energy clustering mechanisms: effective sound-speed suppression, dynamical excitation of the heavy mode, and tachyonic instabilities induced by field-space curvature, and explore their possible imprints on the matter power spectrum and cosmic microwave background anisotropies. Our analysis establishes multifield dark energy as a rich and testable extension of quintessence, with observable clustering signatures relevant to current and future cosmological surveys.
\end{abstract}

\preprint{}
\maketitle


\section{Introduction}
\label{sec:intro}

Understanding the physical origin of the observed late-time cosmic acceleration
\cite{SupernovaCosmologyProject:1998vns,SupernovaSearchTeam:1998fmf}
remains one of the central challenges in modern cosmology.
The phenomenologically simplest explanation is a cosmological constant,
which, together with cold dark matter, defines the concordance $\Lambda$CDM model.
Despite its remarkable empirical success across a wide range of observations
\cite{Planck:2018vyg},
the cosmological constant faces severe theoretical challenges,
most notably the extreme fine-tuning required to reconcile its observed value
with expectations from quantum field theory
\cite{Martin:2012bt,Burgess:2013ara}.

These theoretical shortcomings have motivated extensive investigations of
dynamical dark energy models, in which cosmic acceleration is driven by new
degrees of freedom rather than vacuum energy.
Among these, scalar-field models play a prominent role,
with canonical single-field quintessence
\cite{Copeland:2006wr}
representing the minimal extension beyond $\Lambda$CDM.
In such scenarios, accelerated expansion requires an exceptionally flat scalar
potential, corresponding to a field mass of order the present Hubble scale,
$m \sim H_0$.
As a consequence, single-field dark energy generically behaves as a smooth
component on subhorizon scales, leaving its primary observational imprint on
the background expansion history
\cite{Amendola:2015ksp}.

While single-field constructions are attractive because of their simplicity,
they are difficult to justify from a fundamental perspective.
Low-energy effective theories arising from ultraviolet completions,
most notably string theory, generically predict the presence of multiple
scalar degrees of freedom, such as moduli fields associated with the geometry
of extra dimensions
\cite{Douglas:2006es,Arvanitaki:2009fg}.
From this perspective, single-field quintessence appears as a highly fine-tuned
truncation of a richer multifield theory.
Additional motivation for considering multifield dark energy arises from the
conjectured swampland criteria
\cite{Obied:2018sgi,Agrawal:2018own,Garg:2018reu,Ooguri:2018wrx},
which suggest that scalar potentials in consistent quantum gravity theories
must be sufficiently steep.
These conjectures place single-field quintessence under significant tension
with current cosmological observations
\cite{Akrami:2018ylq,Raveri:2018ddi,Alestas:2024gxe,Akrami:2025zlb},
whereas multifield dynamics can naturally evade these constraints.

A defining feature of multifield scalar theories is the possibility of
nontrivial trajectories in field space.
In particular, when the background evolution follows a strongly non-geodesic
or curved path, accelerated expansion can be sustained even on steep
potentials, in close analogy with the well-known mechanisms of multifield inflation
\cite{Achucarro:2010jv,Achucarro:2012sm,Achucarro:2018vey,Brown:2017osf}.
This decouples the equation of state from the local slope of the potential,
opening new regions of theory space inaccessible to single-field models.
Moreover, curved field-space trajectories have profound implications for
cosmological perturbations, potentially leading to a suppressed sound speed
and enhanced clustering of dark energy on subhorizon scales.

Motivated by these considerations, Ref.~\cite{Akrami:2020zfz}
introduced a class of multifield dark energy models in which cosmic acceleration
proceeds along highly curved trajectories in field space.
Focusing on ``spinning'' solutions with nearly constant angular velocity,
the authors showed that the background cosmology can remain arbitrarily close to
$\Lambda$CDM, while the perturbations exhibit a reduced effective sound speed
and significant clustering on observable scales.
This work demonstrated that multifield dark energy with non-geodesic motion
provides a theoretically well-motivated and phenomenologically rich alternative
to both $\Lambda$CDM and standard quintessence.

Building on this framework, Ref.~\cite{Eskilt:2022hug}
performed a systematic exploration of the background dynamics of spinning
multifield dark energy models with nontrivial field-space geometries.
Through a combination of analytical arguments and numerical phase-space
analyses, the authors established the existence of late-time accelerating
attractor solutions for broad classes of potentials and initial conditions,
while also delineating the regions of parameter space consistent with
swampland constraints and the absence of gradient instabilities.

The present work continues and extends these studies by providing a detailed
analysis of the cosmological perturbations and clustering properties of
spinning multifield dark energy models.
In particular, we investigate the interplay between the exact multifield
description and its effective single-field approximation, clarify the
physical origin of the enhanced growth of structure, and identify distinct
mechanisms through which dark energy perturbations become dynamically
relevant on subhorizon scales.
To this end, we implement the model in \texttt{CAMB} (2021 release)
\cite{Lewis:1999bs},
enabling precise numerical solutions of the complete background and linear
perturbation equations in the synchronous gauge.
This computational framework allows us to trace the evolution of cosmological
quantities, compute matter power spectra, and uncover distinctive signatures
in large-scale structure formation that distinguish this model from both the
concordance $\Lambda$CDM model and single-field quintessence.

Although the multifield models considered here can be practically
indistinguishable from $\Lambda$CDM at the level of the background expansion,
they naturally allow for an evolving dark energy equation of state and can
potentially provide an equally good fit to recent background observations.
This is particularly timely in light of recent results from the
Dark Energy Spectroscopic Instrument (DESI), which point toward dynamical
dark energy when combined with other cosmological probes
\cite{Akrami:2025zlb,Bedroya:2025fwh}.
While background observations alone may not uniquely distinguish such models
from $\Lambda$CDM, their perturbative signatures provide a powerful and
complementary avenue for observational tests.

Throughout this work, we adopt natural units with \(\hbar=c=1\).

\section{Background evolution}
\label{sec:background}
The action for minimally coupled multifield dark energy, characterized by a
field-space metric \(\mathcal{G}_{ab}\), is given by
\begin{equation}\label{action-eq}
S=\int \text{d}^4 x \, \sqrt{-g} \left( \frac{\MPl^2}{2} R - \tfrac{1}{2} \mathcal{G}_{ab} \, \partial_\mu \phi^a \, \partial^\mu \phi^b - V(\phi) + \mathcal{L}_{\text{m}} \right)\,,
\end{equation}
where \(g^{\mu\nu}\) is the spacetime metric, \(\MPl\) is the reduced Planck mass, \(\phi^a\) denotes the scalar fields, \(V\) is the scalar potential,
\(R\) is the Ricci scalar, and \(\mathcal{L}_{\text{m}}\) is the matter
Lagrangian.

Assuming a spatially flat Friedmann--Lema\^{i}tre--Robertson--Walker (FLRW)
background, the Friedmann equation takes the form
\begin{equation}\label{Fridemann-eq}
3 \MPl^2 \mathcal{H}^2 = \tfrac{1}{2} \mathcal{G}_{ab} \, {\phi'}^{a} {\phi'}^{b} + a^2 \, V(\phi) + a^2 \, \rhom \,,
\end{equation}
where \(a\) is the scale factor, \(\mathcal{H}\) is the conformal Hubble parameter, \(\rhom\) is the
matter energy density, and a prime denotes differentiation with respect to
conformal time \(\tau\).

The scalar-field equations of motion are
\begin{equation} \label{fields-eq}
D_\tau {\phi'}^a + 2 \mathcal{H} \phi'^a +  a^2 V^a = 0\,,
\end{equation}
where \(V^a \equiv \frac{\partial}{\partial \phi^a}V\). The operator \(D_\tau\)
denotes the field-space covariant derivative with respect to conformal time
and is defined by
\begin{equation}
    D_\tau A^a \equiv {A^a}' + \Gamma^a_{bc} A^b \phi'^c,
\end{equation}
where \(\Gamma^a_{bc}\) are the Christoffel symbols associated with the
field-space metric.

We focus on a two-field model with \(\phi^a=(r,\theta)\) and adopt the
power-law field-space metric introduced in Ref.~\cite{Eskilt:2022hug},
\begin{equation} \label{metric-eq}
    \text{d}s^2 = \text{d}r^2 + f\,\text{d}\theta^2\,,
\end{equation}
where \(f=f(r)=r^p\). The corresponding field-space Ricci scalar is
\begin{equation}
    \mathcal{R} = \frac{p\,(2-p)}{2 \, r^2}\,.
\end{equation}
For \(p=2\), the metric reduces to the flat polar-coordinate metric.
Values of \(p>2\) correspond to negatively curved field spaces, whereas
\(0<p<2\) yields positive field-space curvature.

This metric provides a simple framework for investigating the effects of
non-geodesic scalar-field dynamics and field-space curvature on the
cosmological evolution.

Since the power-law metric becomes singular at \(r\le0\), it may lead to
numerical instabilities in both the background and perturbation equations.
To avoid these singularities, we enforce \(r>0\) throughout the numerical
evolution.

Using Eqs.~(\ref{Fridemann-eq}) and (\ref{fields-eq}), the background
equations of motion for the metric~(\ref{metric-eq}) are
\begin{align}
& 3 \MPl^2 \cH^2 = \half \left( r'^2 + f \, \theta'^2 \right) + a^2 \, V + a^2 \rho_m\,, \label{be1_eq}\\
& r'' + 2 \cH r' -\frac{1}{2} f_r \, \theta'^2 + a^2 V_r = 0\,, \label{be2_eq}\\
& \theta'' + 2 \cH \theta' + \frac{f_r}{f}{ r' \, \theta'} + a^2 \, \frac{1}{f} \,V_\theta = 0\,, \label{be3_eq}
\end{align}
where the subscripts \(r\) and \(\theta\) denote partial derivatives with
respect to \(r\) and \(\theta\), respectively.

We adopt the potential introduced in the previous studies
\cite{Eskilt:2022hug,Akrami:2020zfz}:
\begin{equation}
V(r, \theta) = V_0 - \alpha \theta + \frac{1}{2} m^2 (r - r_0)^2\,,
\end{equation}
where \(V_0\), \(\alpha\), \(m\), and \(r_0\) are free parameters to be constrained by observational data.
As discussed in Ref.~\cite{Akrami:2020zfz}, this potential provides a simple
effective field-theory construction that supports the non-geodesic field-space
trajectories considered in this work. It consists of a quadratic radial potential and a linear angular term that softly breaks the shift symmetry in \(\theta\), thereby giving rise to the desired dynamical behavior.

Ref.~\cite{Eskilt:2022hug} derived an approximate analytical solution that elucidates the dynamics of the scalar fields during the cosmological background evolution.
In the late-time attractor, under nearly complete dark energy domination (\(\Omega_\phi \simeq 1\)) and within the potential-dominated slow-roll regime,
the radial potential gradient is balanced by the geometric (centrifugal) effect generated by the angular motion,
\(a^2 V_r \simeq f_r \theta'^2/2\),
keeping \(r\) close to a quasi-equilibrium value \(r_{\rm eq}\).
As \(\theta\) slowly rolls down its potential, the equilibrium position \(r_{\rm eq}\) shifts accordingly,
so that the two fields evolve in a coupled manner.
The algebraic equation determining \(r_{\rm eq}\) can be approximated by
\begin{equation}
\frac{V_r }{V_\theta^2} \simeq \frac{3\MPl^2}{8 V}\,\frac{f_r}{f^2}\,,
\end{equation}
which follows from Eqs.~(\ref{be1_eq})--(\ref{be3_eq}) under the assumptions that \(r''\), \(r'\), and \(\theta''\) are small compared to the remaining terms
\cite{Eskilt:2022hug}.

Physically, the system follows a non-geodesic (turning) trajectory in field
space, where the angular motion continuously maintains the equilibrium value
of \(r\). This allows the Universe to accelerate even for steep potentials while keeping \(w_\phi \approx -1\) at late times.
The agreement between the approximate and numerical values of \(r_{\rm eq}\) has been investigated in Ref.~\cite{Eskilt:2022hug}, to which we refer the reader for further details.

In the remainder of this paper, we work in reduced Planck units, setting
\(M_{\rm Pl} \equiv (8\pi G)^{-1/2} = 1\). The scalar fields \(r\) and
\(\theta\), together with their perturbations, are dimensionless.
Conformal-time derivatives carry the appropriate mass dimension, and all
dimensionful quantities are measured in powers of \(M_{\rm Pl}\).

We solve the background equations numerically in conformal time.
Although the background evolution depends only weakly on the initial conditions
\(r_i\), \(r'_i\), and \(\theta'_i\),
obtaining accurate numerical solutions for the perturbation equations and
performing parameter estimation require us to constrain the evolution of the
quintessence fields so that the present epoch of the numerical evolution
coincides with that of the observed Universe.
This ensures that the total density parameter,
\(\Omega_{\rm tot}=\rho_{\rm tot}/\rho_{\rm crit}\),
is equal to unity at redshift \(z=0\).

\begin{figure}[!]
    \noindent
    \includegraphics[width=1\columnwidth]{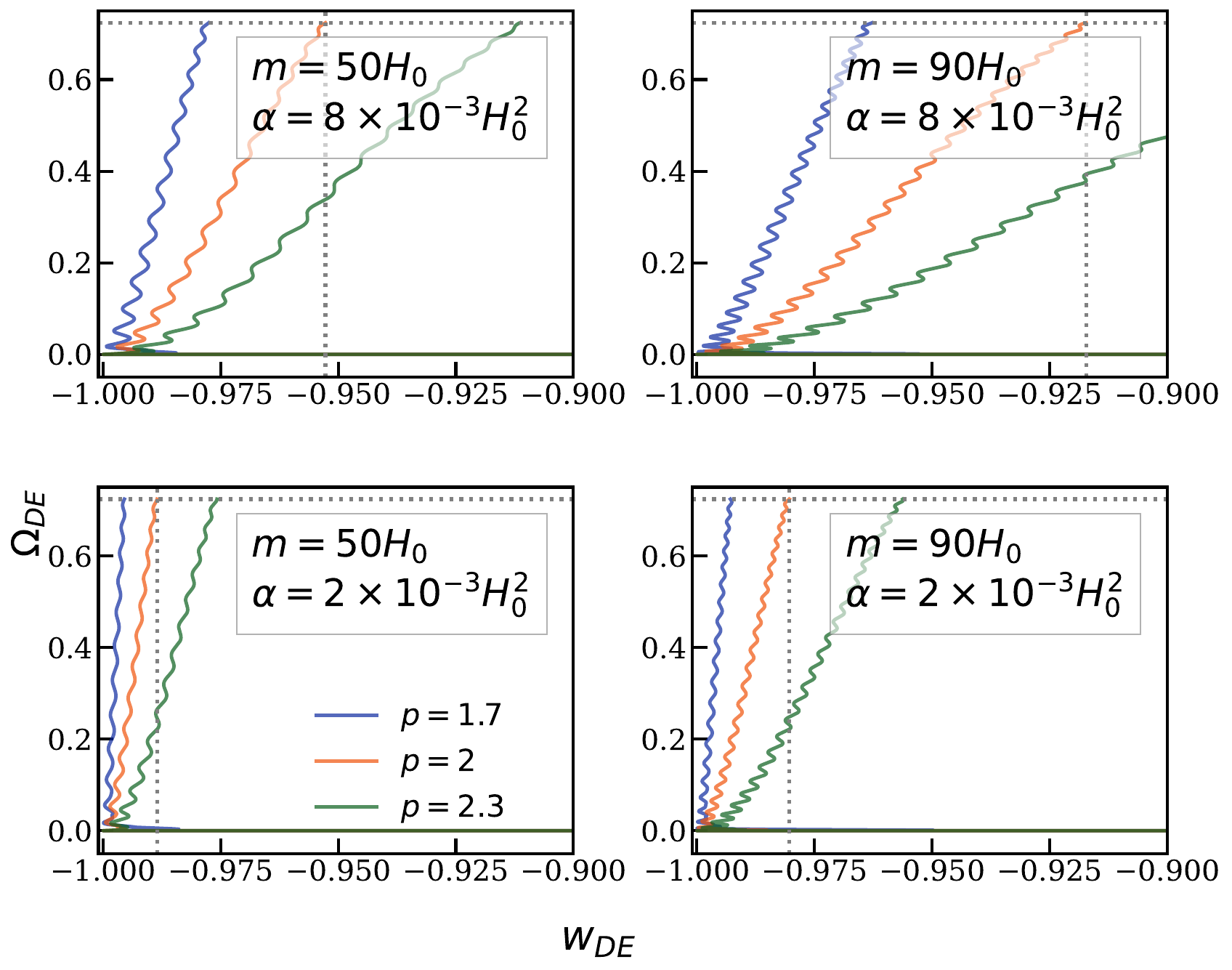}
    \caption{Phase-space evolution of the fractional dark energy density parameter
\(\Omega_{\rm DE}\) as a function of the dark energy equation-of-state parameter
\(w_{\rm DE}\) for different values of the field-space curvature parameter
\(p\), \(m\) (in units of \(H_0\)), and \(\alpha\) (in units of
\(H_0^2\)). The initial conditions are fixed to
\(r_i=7\times10^{-4}\), \(r'_i=0.1\), and \(\theta'_i=100\), with
\(r_0=7\times10^{-4}\).
The vertical dotted line indicates the value of \(w_{\rm DE}\) at the
epoch when \(\Omega_{\rm DE}\) reaches its present-day value, obtained by
imposing \(\Omega_{\rm tot}=1\) at \(z=0\). The case \(p=2\) is shown as a
representative example.}
    \label{fig:omega_vs_w}
\end{figure}

Figure~\ref{fig:omega_vs_w} illustrates the phase-space evolution of the
fractional dark energy density parameter \(\Omega_{\rm DE}\) as a function of
the dark energy equation-of-state parameter \(w_{\rm DE}\) for different
values of \(p\), corresponding to different field-space curvatures.
The initial conditions are identical in all cases,
\(r_i=7\times10^{-4}\),
\(r'_i=0.1\),
and \(\theta'_i=100\).
In addition, the vacuum expectation value \(r_0\) is fixed to
\(7\times10^{-4}\) throughout.

The oscillatory behavior of \(\Omega_{\rm DE}\) arises from the nonzero initial
velocities of the \(r\) and \(\theta\) fields.
These conditions cause the field trajectory to deviate from a circular orbit
as \(r\) climbs the potential.
Such oscillations are particularly pronounced when the initial field
configuration is significantly displaced from \(r=r_0\).

In this figure, we examine how the parameters \(m\) and \(\alpha\) affect the
evolution of the equation of state up to the present epoch.
The parameter \(V_0\) is determined by imposing the present-day total density
constraint.
While it is desirable for the model to admit late-time accelerating attractors
over a broad range of initial conditions, the weak dependence of the present-day
equation of state on those initial conditions improves the model's ability to
remain consistent with recent observational data.

For instance, for \(m=90H_0\), \(\alpha=8\times10^{-3}H_0^2\), and \(p=2\),
where \(H_0\) denotes the present-day Hubble expansion rate, the dark-energy
equation-of-state parameter is approximately \(w_{\rm DE}\approx-0.92\).
This value is in good agreement with recent analyses of the DESI Data Release~2
(DR2) baryon acoustic oscillation measurements
\cite{Akrami:2025zlb}.
The authors of Ref.~\cite{Akrami:2025zlb} have shown that a single-field
exponential quintessence model with potential
\(V(\phi)=V_0e^{-\lambda\phi}\) may provide a better fit than the standard
\(\Lambda\)CDM model to the combined DESI DR2, {\it Planck} cosmic microwave
background (CMB) distance priors, and Dark Energy Survey Year~5 (DES Y5)
supernova datasets, depending on the statistical inference measure adopted and
on whether the spatial curvature parameter \(\Omega_K\) is allowed to vary.
The best-fit value, \(\lambda\approx0.6\) (68\% confidence level),
corresponds to a present-day equation-of-state parameter
\(w_{\rm DE}\approx-0.92\), which may be interpreted as supporting the
viability of dynamical dark energy models such as ours in accommodating the
evolving dark-energy equation of state suggested by the DESI data.

The numerical results are consistent with the approximate analytical
expression for the equilibrium radius \(r_{\rm eq}\) derived in
Ref.~\cite{Eskilt:2022hug}, which yields the following expression for the
dark energy equation-of-state parameter:
\begingroup
\setlength{\abovedisplayskip}{6pt}
\setlength{\belowdisplayskip}{6pt}
\setlength{\abovedisplayshortskip}{6pt}
\setlength{\belowdisplayshortskip}{6pt}
\begin{equation}\label{EoS_eq}
w_{\rm DE} = -1 + \dfrac{2}{1 + p\!\left(\dfrac{1}{2} + \dfrac{V_0 - \alpha\,\theta}{m^2 r_{\rm eq}^2}\right)}\,.
\end{equation}
\endgroup
Increasing either \(m\) or \(\alpha\) shifts \(w_{\rm DE}\) away from \(-1\)
toward larger values.

As discussed above, the choice of initial conditions can induce oscillatory
behavior at early times.
However, the late-time background evolution is largely insensitive to these
initial conditions.
As indicated by Eq.~(\ref{EoS_eq}), the model's ability to fit observational
data is determined primarily by the parameters of the potential.

As we show in the following sections, the initial conditions play a crucial
role in the evolution of perturbations and in shaping the clustering behavior
of dark energy.
Consequently, the model can be constrained using observations of both the
background evolution and the perturbation dynamics.

\begin{figure}[H]
    \noindent
    \includegraphics[width=1\columnwidth]{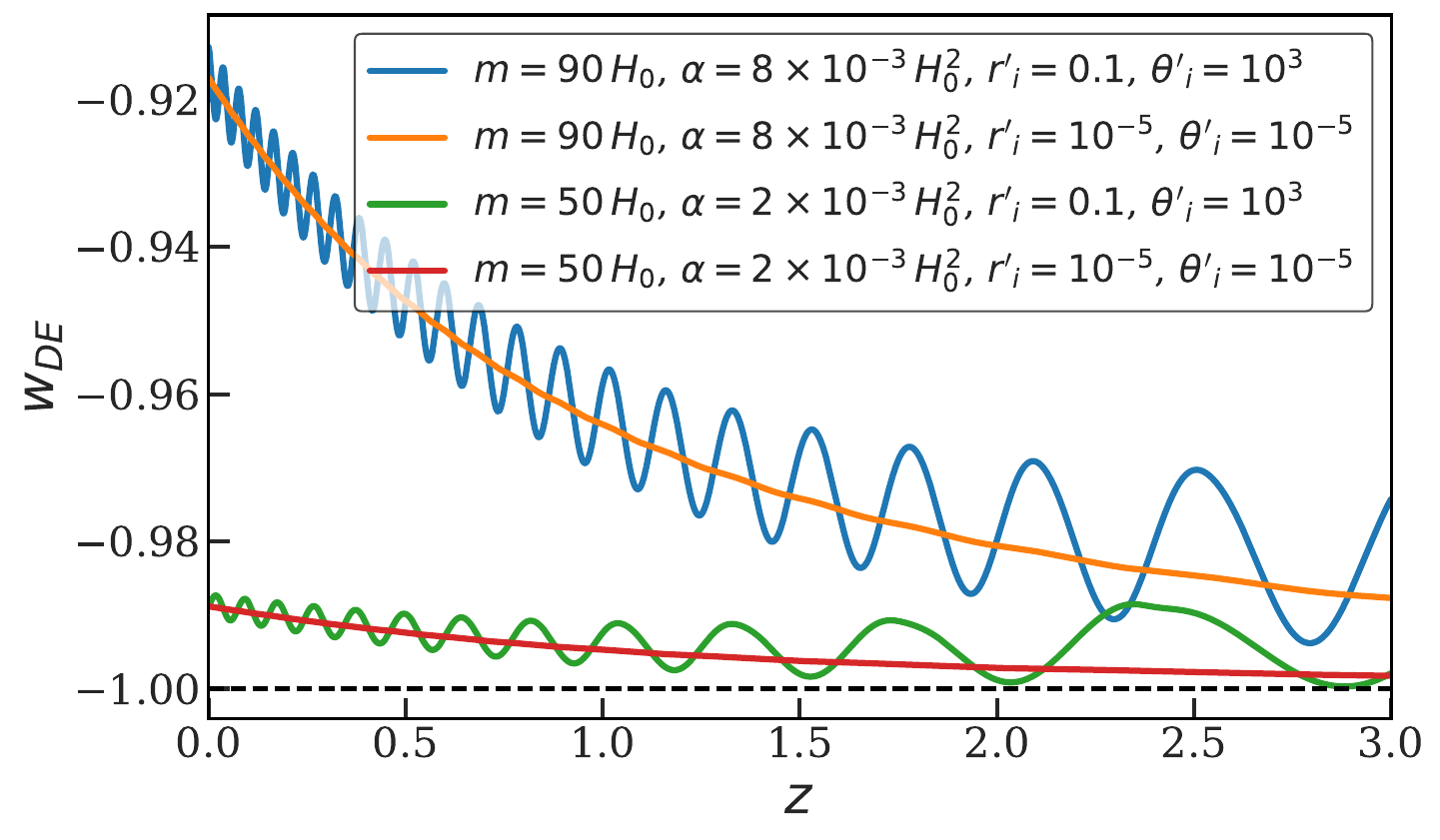}
    \caption{Variation of the dark-energy equation-of-state parameter \(w_{\rm DE}\) as a function
of redshift \(z\) for different parameter sets and initial conditions, with
\(p=2\) and \(r_i=r_0=7\times10^{-4}\).}
    \label{fig:w_vs_z}
\end{figure}

Figure~\ref{fig:w_vs_z} simultaneously illustrates the effects of the initial
conditions and the potential parameters on the background evolution.

For a given set of potential parameters, the figure shows that larger initial
field velocities corresponding to hard initial
conditions\,\footnote{Hard initial conditions refer to large initial
scalar-field velocities for which the field trajectory undergoes oscillations
and the effective single-field description breaks down at late times, whereas
soft initial conditions correspond to small initial velocities for which the
field trajectory smoothly relaxes toward a dynamical equilibrium in field
space and can be accurately described by an effective single-field theory.}
do not significantly modify the present-day value of the equation-of-state
parameter.

This demonstrates that the background evolution is relatively insensitive to
the choice of initial conditions.
Oscillations induced by large initial velocities are progressively damped by
cosmic expansion and become negligible at late times.

Our background evolution exhibits no evidence of phantom behavior over a wide
range of parameter choices and initial conditions.

\section{Linear perturbations and mode decomposition}\label{sec:Perturbation}
In this study, we use the synchronous-gauge formulation for the perturbation
equations of a general multifield dark energy model. The line element in the
synchronous gauge is
\begin{equation}
    \text{d}s^2 = a^2(\tau) \left[ - \text{d}\tau^2 + (\delta_{ij} + h_{ij})\, \text{d}x^i \text{d}x^j \right],
\end{equation}
where \(a(\tau)\) is the scale factor and \(h_{ij}\) denotes the linear
perturbation of the spatial metric \cite{1995ApJ...455....7M}.
After perturbing the scalar-field equations of motion and using the background
equations~(\ref{fields-eq}), the scalar-field perturbations in Fourier space can
be written as \cite{Akrami:2020zfz,Eskilt:2022hug}
\begin{align} \label{pert_eq}
D^2_\tau \delta \phi^c + 2 \cH D_\tau \delta \phi^c + k^2 \delta \phi^c
- \mathcal{R}^c{}_{abd}\, \phi^{a'} \phi^{b'} \delta \phi^d  \nonumber\\
+ \frac{1}{2}h'\, \phi^{c'} + a^2 D_e D^c V\, \delta \phi^e = 0\,,
\end{align}
where \(\mathcal{R}^c{}_{abd}\) is the field-space Riemann tensor,
\(h \equiv \delta^{ij} h_{ij}\) is the trace of the spatial metric
perturbation,\footnote{The symbol \(h\) used here for the trace of the metric perturbation
should not be confused with the dimensionless Hubble parameter \(h\) used
elsewhere in this paper.} and \(D_e\) and \(D^c\) denote the field-space covariant
derivatives associated with the metric \(\mathcal{G}_{ab}\).

It is useful, both analytically and numerically, to decompose the perturbations
along the normalized tangent vector \(\boldsymbol{\Ta}\) and the corresponding
normal vector \(\boldsymbol{\N}\), defined by
\begin{equation}\label{TN_eq}
    \Ta^a \equiv \frac{\phi'^a}{\phi'}\,, \qquad
      \N^a \equiv - \frac{1}{a\Omega} D_\tau \Ta^a\,.
\end{equation}
Here, \(\phi'^2 = \G_{ab} \phi'^a \phi'^b\), and
\(a\Omega = |D_\tau \Ta|\) defines the turning rate, which measures the
deviation from a geodesic trajectory. For a two-dimensional scalar-field space,
\(\Omega\) can be written as
\begin{equation}\label{Omega_eq}
     \Omega = \frac{a}{\sqrt{\text{det} \G}} \frac{|\phi'_1 V_2 - \phi'_2 V_1|}{\phi'{}^2}\,.
\end{equation}

The scalar-field equations of motion~(\ref{fields-eq}) projected along these
directions become
\begin{align}\label{TN_field_eq}
    &\phi'' + 2 \mathcal{H} \phi' + a^2 D_\phi V = 0\,, \nonumber \\
    &\Omega\, \phi'  - a D_\mathcal{N} V = 0\,.
\end{align}
Here, \(D_\phi \equiv \Ta^a D_a = (1/\phi') D_\tau \) and
\(D_\N \equiv \N^a D_a\).

Starting from Eq.~(\ref{pert_eq}) and using
Eqs.~(\ref{TN_eq}) and~(\ref{TN_field_eq}), the perturbation equations in the
tangent--normal basis become
\begin{align}
\delta \phi^{\Ta \prime \prime} &+ 2 \cH \delta \phi^{\Ta \prime}
 + 2 a \Omega \delta \phi^{\N \prime} 
 + \left( k^2 + a^2 D_\phi^2 V \right) \delta \phi^\Ta \nonumber\\
 & + 2 a \Omega \!\left(2 \cH + \frac{\Omega'}{\Omega}
   +\frac{\phi''}{\phi'} \right) \!\delta \phi^\N 
 + \frac{1}{2} h' \phi^{\prime} = 0\,, \label{delphiT_eq}\\
\delta \phi^{\N \prime \prime} &+ 2 \cH \delta \phi^{\N \prime}
 - 2 a \Omega \delta \phi^{\Ta \prime} 
 + \left( k^2 + \mathcal{M}_{\rm{eff}}^2 \right) \delta \phi^\N  \nonumber\\
 &- 2 a \Omega \!\left(\cH - \frac{\phi''}{\phi'} \right) 
   \delta \phi^\Ta = 0\,. \label{delphiN_eq}
\end{align}
In these equations, \(\delta \phi^\Ta = \Ta_a \delta \phi^a\) and
\(\delta \phi^\N = \N_a \delta \phi^a\). The effective mass of the normal
perturbation \(\delta\phi^\N\) is
\begin{equation}
    \mathcal{M}_{\rm{eff}}^2 = a^2 V_{\N\N} - a^2 \Omega^2 + \mathcal{R} \frac{\phi'^2}{2}\,.
\end{equation}
Here, \(V_{\N\N} = \N^a \N^b D_a D_b V\).

The linearized Einstein equations in \(k\)-space are
\begin{align}
&k^2 \eta - \half \cH h' = \frac{a^2}{2 \MPl^2}\;(\rho_{\rm{DM}} \delta_{\rm{DM}} + \delta\rhoDE)\,, \label{delrho_eq}\\
&k^2 \eta' =  \frac{a^2}{2 \MPl^2}(\rhoDE+\PDE)\Theta\,,
\end{align}
where \(\rho_{\rm DM}\) is the background dark matter density, \(h\) is the
trace of the metric perturbation \(h_{ij}\) defined above, and \(\eta\) is the
scalar potential that parameterizes the traceless part of the metric
perturbation in Fourier space in the synchronous gauge
\cite{1995ApJ...455....7M}. Furthermore, \(\delta\rho_{\rm DE}\) is the
dark-energy density perturbation, and \(\Theta\) is the corresponding velocity
divergence of the scalar field in this gauge. These quantities are given by
\begin{align}
&\delta \rho_{\text{DE}} = 
\frac{\phi^{\prime 2}}{a^2} \!\left( 
 \frac{\delta \phi^\Ta}{\phi'} + \frac{a \Omega \delta \phi^\N}{\phi'} \right)
 + D_\phi V \delta \phi^\Ta + D_\N V \delta \phi^\N\,,\\
&(\rhoDE+\PDE) \Theta  = \frac{k^2}{a^2} \phi' \delta \phi^\Ta\,. \label{delq_eq}
\end{align}

For the numerical analysis, we solve Eqs.~(\ref{delphiT_eq})--(\ref{delq_eq})
together with the background equations~(\ref{be1_eq})--(\ref{be3_eq}) directly
within the \texttt{CAMB} (2021) framework.

In the following section, we compare the full numerical results with an
analytical interpretation of the model.


\section{Comparison between exact and effective perturbation descriptions}
\label{EFT-exact}

In the following, we consider the sub-horizon limit of the perturbation equations to verify
the consistency between the effective analytical description and the full numerical solutions.
Although our formulation is developed in the synchronous gauge, the resulting reduced perturbation equations
are equivalent to the sub-horizon limit of the Newtonian-gauge perturbation equations
derived in Ref.~\cite{Akrami:2020zfz}:

\begin{align} 
&\delta \phi^{\Ta \prime \prime} + 2 a \Omega \delta \phi^{\N \prime} 
+ k^2 \delta \phi^\Ta = 0\,, \label{modified1:eq}\\
&\delta \phi^{\N \prime \prime} - 2 a \Omega \delta \phi^{\Ta \prime}
+ ( k^2 + \mathcal{M}_{\rm{eff}}^2 ) \delta \phi^\N = 0\,. \label{modified2:eq}
\end{align}

These equations are obtained under the sub-horizon assumption
\(k^2 \gg \mathcal{H}^2\), while neglecting gravitational backreaction,
matter perturbations, and the light-mode mass term
\(a^2 D_\phi^2 V\) in the tangential equation.

The corresponding dispersion relation yields a modified sound speed
\(c_\text{s}\) for the light mode,
\(\omega_-^2=c_\text{s}^2k^2\), given by
\begin{equation} \label{eq:cs2sm}
  \frac{1}{c_\text{s}^2} = 1 + \frac{4 a^2 \Omega^2}{\mathcal{M}_{\rm{eff}}^2}\,,
\end{equation}
where the approximation assumes
\(k^2 \ll \mathcal{M}_{\rm{eff}}^2 + 4a^2\Omega^2\).
As discussed in Ref.~\cite{Akrami:2020zfz}, the sound speed is suppressed when
\(a^2\Omega^2 \gg \mathcal{M}_{\rm{eff}}^2\), leading to enhanced clustering on
sub-horizon scales.

For the heavy mode, the dispersion relation to quadratic order in \(k\) is
\begin{equation}
    \omega_+^2 = \frac{\mathcal{M}_{\rm{eff}}^2}{c_{\rm s}^2}
    + (2-c_\text{s}^2)k^2 + \mathcal{O}(k^4)\,.
\end{equation}
Moreover, owing to the suppressed propagation speed of the light mode,
the heavy mode can be consistently integrated out over a broad range of
intermediate scales,
\begin{equation} \label{heavyMode_ineq}
\mathcal{H}^2 \ll k^2 \ll \frac{\mathcal{M}_{\rm{eff}}^2}{c_\text{s}^2}\,.
\end{equation}
In this regime, the system admits an effective single-field description,
valid within the sub-horizon approximation, in which the physical effects of
the heavy mode are encoded in the modified dispersion relation of the light
mode.

In the synchronous gauge, the sound speed of field fluctuations is
formally defined as
\(c_{\text{s};\rm synch}^2\equiv\delta\PDE/\delta\rhoDE\),
where \(\delta\PDE\) denotes the dark energy pressure perturbation in the
synchronous gauge.
It is important to emphasize, however, that the definition of the sound
speed is not universal and is, in general, gauge dependent.
Therefore, assessing the stability of a cosmological model directly from the
perturbation equations requires transforming the perturbations to the
dark-energy rest frame, where unphysical gauge artifacts are removed and the
physical propagation speed of the perturbations can be defined
unambiguously.

The gauge transformations relating the dark-energy rest-frame density
and pressure perturbations to their synchronous-gauge counterparts are
given by \cite{PhysRevD.69.083503},
\begin{align}
    \delta\hat{\rho}_\mathrm{DE} &= \delta\rhoDE - \frac{\Theta}{k^2}\,\rhoDE'\,,\\
    \delta\hat{P}_\mathrm{DE} &= \delta \PDE - \frac{\Theta}{k^2}\,\PDE'\,,
\end{align}
where the hat denotes quantities evaluated in the dark-energy rest frame.
The physical sound speed of dark-energy perturbations is therefore defined as
\begin{equation}
    \hat{c}_s^2 = \frac{\delta \hat{P}_\mathrm{DE}}{\delta \hat{\rho}_\mathrm{DE}}\,.
\end{equation}

\begin{figure}[!]
    \noindent
    \includegraphics[width=1\columnwidth]{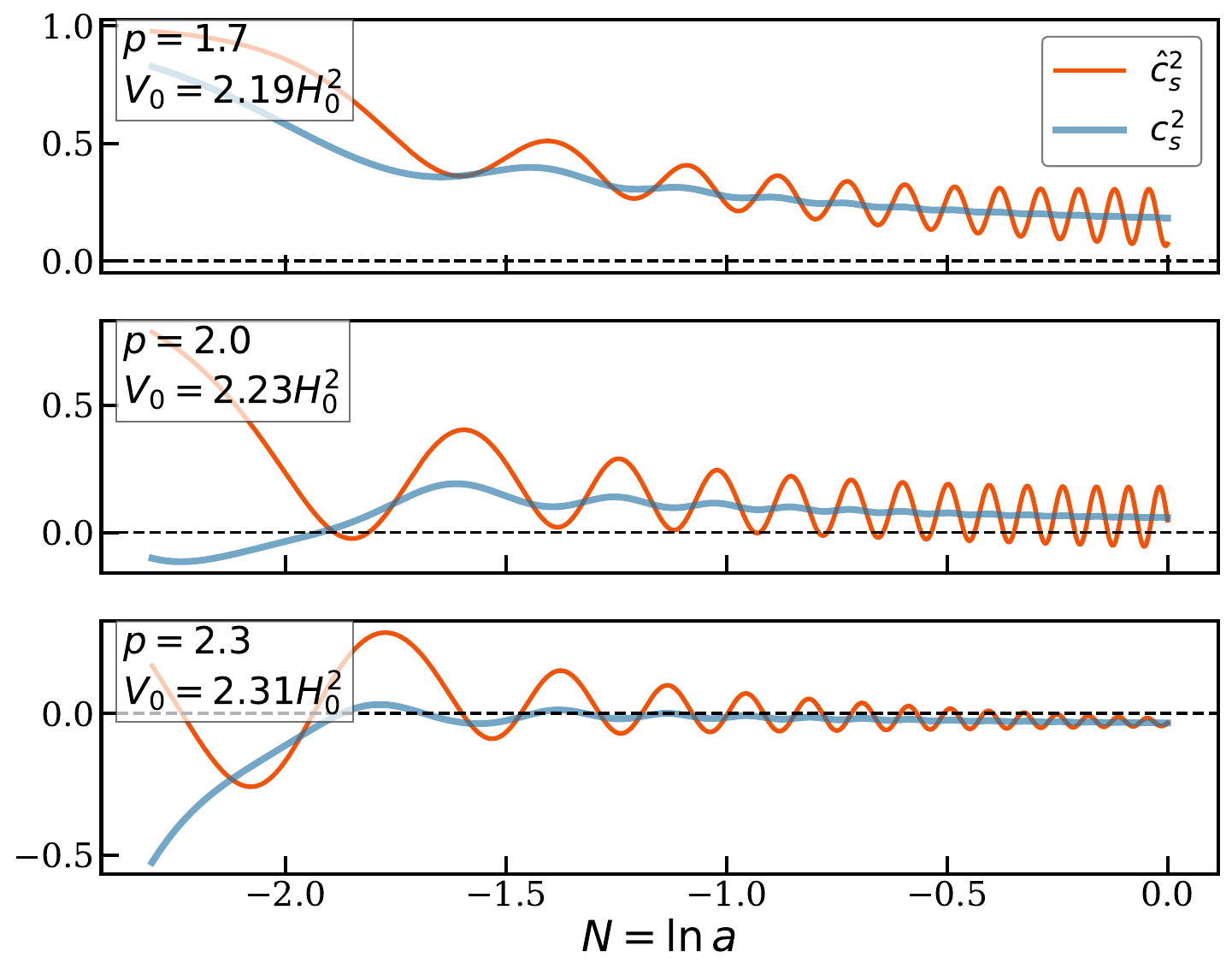}
\caption{
Evolution of the rest-frame sound speed \(\hat{c}_s^2\) and the modified
sound speed \(c_\text{s}^2\) as functions of the $e$-fold number \(N=\ln a\)
for different values of the parameter \(p\) and the corresponding
potential amplitudes \(V_0\) (indicated in each panel).
The results are obtained for a comoving wavenumber \(k=0.001 \,h\,\mathrm{Mpc}^{-1}\),
with model parameters \(\alpha=0.002H_0^2\),
\(m=50H_0\),
\(r_0=r_i=7\times10^{-4}\),
\(r'_i=10^{-5}\), and
\(\theta'_i=10^{-5}\).
}
\label{fig:cs2}
\end{figure}

The excellent agreement between the rest-frame sound speed
\(\hat{c}_s^2\) and the modified sound speed \(c_\text{s}^2\)
is illustrated in Fig.~\ref{fig:cs2}.
The panels show their evolution as functions of the $e$-fold number
\(N=\ln a\) for different values of the parameter \(p\),
namely \(p=1.7\), \(2.0\), and \(2.3\), corresponding to
positive, zero, and negative field-space curvature, respectively.
The results are computed for the comoving wavenumber \(k=0.001 \,h\,\mathrm{Mpc}^{-1}\),
which lies within the sub-horizon regime where dark-energy perturbations are
dynamical and clustering may occur, depending on the initial conditions.
We assume that the scalar-field evolution begins during a matter-dominated
epoch with vanishing perturbations.

The orange curves represent the rest-frame sound speed
\(\hat{c}_s^2\), computed numerically from the full perturbation
equations, while the blue curves show the modified sound speed
\(c_\text{s}^2\) of the light mode obtained from the reduced
equations~(\ref{modified1:eq}) and~(\ref{modified2:eq}).
The oscillatory behavior of \(\hat{c}_s^2\) originates from the full
two-field dynamics and the time dependence of the background trajectory.
Despite the difference in amplitude, the oscillation phase of
\(\hat{c}_s^2\) closely follows that of \(c_\text{s}^2\), indicating that both
quantities are governed by the same underlying light-mode dynamics on
sub-horizon scales.

The reduced amplitude of the modified sound speed relative to the rest-frame sound
speed reflects the assumptions used in deriving the reduced perturbation equations,
namely the neglect of gravitational backreaction and the assumption that the
background evolution has already reached a slowly varying turning regime, in which
\(\phi'' \ll \mathcal{H}\phi'\) and \(\Omega' \ll \mathcal{H}\Omega\).
In this regime, both the turning rate and \(\phi'\) evolve adiabatically on
cosmological timescales, and the scalar-field trajectory approaches a turning
attractor characterized by a nearly constant turning rate around
\(r=r_{\rm eq}\).

In the exact perturbation equations~(\ref{delphiT_eq}) and~(\ref{delphiN_eq}),
however, these conditions are generally not satisfied over a broad range of
parameter values and initial conditions.
In particular, the finite age of the Universe prevents the background trajectory
from fully relaxing to the quasi-equilibrium configuration \(r=r_{\rm eq}\) by
the present epoch (\(z=0\)).
As a result, the rest-frame sound speed exhibits persistent deviations from the
effective description.

More precisely, although the high- and low-energy perturbation eigenmodes
effectively decouple, the larger oscillation amplitude arises from the incomplete
relaxation of the background trajectory.
Consequently, residual effects of the heavy field can leave non-negligible
imprints on the evolution of the light mode beyond the simple suppression of the
sound speed~\cite{Xian_Gao_2012,Achucarro:2012sm}.%
\footnote{Throughout this paper, the \emph{heavy field} and \emph{light field}
refer to the normal and tangential field-space directions, respectively,
whereas the \emph{heavy mode} and \emph{light mode} denote the high- and
low-energy eigenmodes of the perturbation system, corresponding to the
frequencies \(\omega_+\) and \(\omega_-\)
\cite{Achucarro:2012sm}.}

For \(p=1.7\) (top panel), both \(\hat{c}_s^2\) and \(c_\text{s}^2\) are positive at
early times (large negative \(N\)) and remain positive at late times, with no
significant suppression, indicating stable perturbations without growing
instabilities.
For \(p=2.0\) (middle panel), \(\hat{c}_s^2\) decreases at late times from
positive values toward \(c_\text{s}^2\) and oscillates around it, with brief negative
excursions before returning to positive values owing to the oscillatory
dynamics.
For \(p=2.3\) (bottom panel), both \(\hat{c}_s^2\) and \(c_\text{s}^2\) evolve toward
negative values at late times, indicating the possibility of growing gradient
instabilities.\footnote{A negative modified sound speed squared in the effective single-field
description of this model does not correspond to a genuine gradient
instability, but instead reflects a (possibly transient) tachyonic instability
of the heavy field projected onto the light sector. The associated growth is
regulated by spatial gradient terms and is therefore limited to a finite range
of scales~\cite{Garcia-Saenz_2018}.}

To demonstrate the excellent consistency between the full numerical solution of
the perturbation equations and the analytical description derived from the
modified equations, it is useful to consider the scale-dependent sound speed
associated with the light-mode branch,
\(c_{{\rm s}k}\equiv c_\text{s}(k)\), obtained from the modified dispersion relation of the
light mode to quadratic order in \(k\).
This quantity interpolates between the low-energy light-mode behavior and the
relativistic regime at large wavenumbers and is given by~\cite{Xian_Gao_2012,PhysRevD.86.121301}
\begin{equation} \label{cs2effk_eq}
    c_{{\rm s}k}^2 = \frac{\mathcal{M}_{\rm{eff}}^2 + k^2}
{\mathcal{M}_{\rm{eff}}^2/c_\text{s}^2 +  k^2 }\,.
\end{equation}

\begin{figure}[!h]
    \noindent
    \includegraphics[width=\columnwidth]{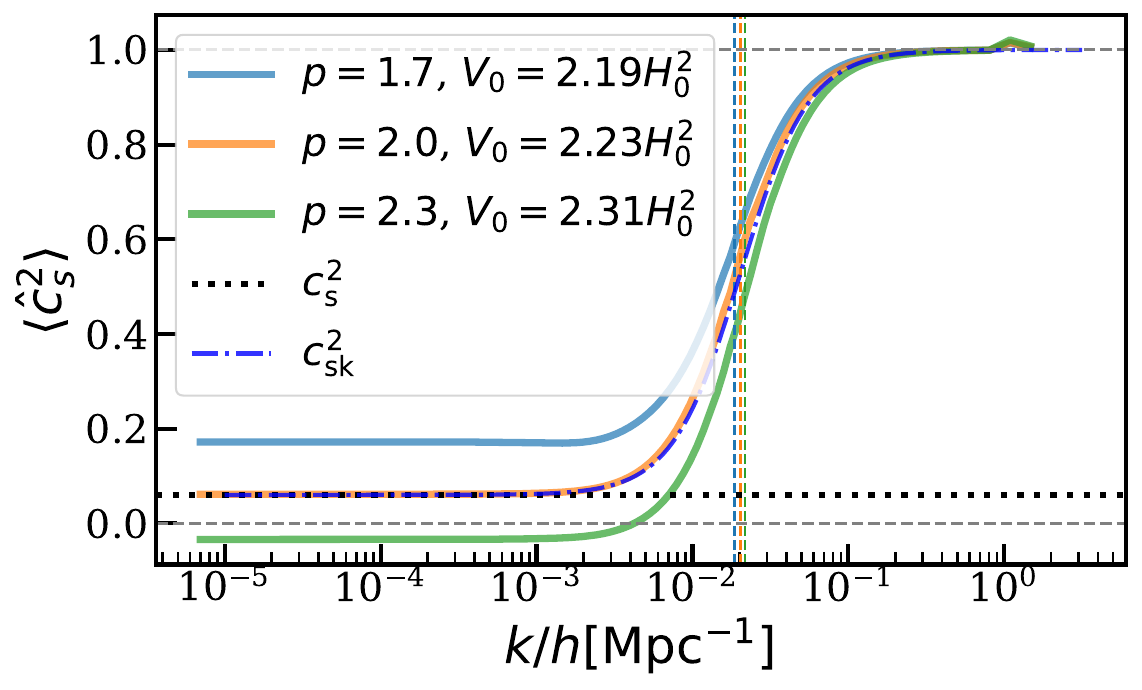}
    \caption{
    The averaged rest-frame sound speed squared,
    \(\langle \hat{c}_s^2 \rangle\), shown as a function of the comoving
    wavenumber \(k\).
    The average is computed over the interval \(a\in(0.8,1)\) for three
    different values of the parameter \(p\).
    The vertical dashed lines indicate the validity limit of the effective
    single-field description, given by
    \(\mathcal{M}_{\rm eff}/c_\text{s}\), evaluated at \(a\approx1\) for \(p=2\).
    The blue dash--dot curve shows the scale-dependent sound speed
    \(c_{{\rm s}k}\), and the horizontal black dotted line corresponds to the
    modified light-mode sound speed evaluated at \(a\approx1\) and \(p=2\).
    The remaining parameter values are the same as in
    Fig.~\ref{fig:cs2}:
    \(\alpha=0.002H_0^2\), \(m=50H_0\), \(r_0=7\times10^{-4}\),
    \(r_i=7\times10^{-4}\), \(r'_i=10^{-5}\), and
    \(\theta'_i=10^{-5}\).
    }
    \label{fig:cs2_k}
\end{figure}

Figure~\ref{fig:cs2_k} shows the averaged rest-frame sound speed squared,
\(\langle \hat{c}_s^2 \rangle\), computed over the interval
\(a\in(0.8,1)\) (solid curves) from the full numerical solution of the
perturbation equations~(\ref{delphiT_eq}) and~(\ref{delphiN_eq}),
together with the modified scale-dependent sound speed
\(c_{{\rm s}k}\) (blue dash--dot curve), computed from the background evolution
for \(p=2\).

Depending on the value of the field-space curvature parameter \(p\),
the averaged rest-frame sound speed is strongly suppressed at low
wavenumbers, approaching zero or even negative values. This indicates that
the normal and tangential field-space components remain coupled, while the
heavy mode can be consistently integrated out within the effective
description.
In contrast, at sufficiently large wavenumbers, the sound speed approaches
unity, corresponding to relativistic propagation in which the scalar-field
components become effectively decoupled and clustering on small
sub-horizon scales is suppressed.

For \(p=2\), the agreement between
\(\langle \hat{c}_s^2 \rangle\) and \(c_{{\rm s}k}^2\) is clear.
This excellent consistency between the effective description and the exact
perturbation evolution results from adopting soft initial conditions, under
which the high-energy eigenmode is not significantly excited and the
oscillatory behavior of the rest-frame sound speed is efficiently averaged
out \cite{PhysRevD.86.121301}.

The vertical dashed lines indicate the upper bound,
\(\mathcal{M}_{\mathrm{eff}}/c_\text{s}\),
above which the heavy mode can no longer be integrated out and the system can
no longer be accurately described by an effective single-field theory.
As expected, the suppression of the averaged rest-frame sound speed sets in
at wavenumbers well below the locations of the vertical dashed lines.
This is consistent with the validity condition of the effective single-field
theory, which requires
\begin{equation}
k^2 \ll \frac{\mathcal{M}_{\mathrm{eff}}^2}{c_\text{s}^2}\,.
\end{equation}

The asymptotic behavior of the modified sound speed in the low-\(k\) limit at
late times is given by \cite{Eskilt:2022hug}
\begin{equation}
c_\text{s}^2 \simeq \frac{2-p}{2+p}\,.
\end{equation}
This expression directly illustrates how the field-space curvature affects
the sound speed.
As shown in Fig.~\ref{fig:cs2_k}, the averaged rest-frame sound speed
obtained from the full numerical solution follows this analytical prediction
at sufficiently small wavenumbers, approaching the corresponding asymptotic
light-mode value as \(k\to0\).
This agreement demonstrates that, in the long-wavelength regime, the full
perturbation dynamics reproduces the expected asymptotic light-mode behavior
encoded in the modified dispersion relation.
Larger values of \(p\) lead to stronger suppression and can render
\(c_\text{s}^2\) negative, whereas smaller values of \(p\) produce a weaker
suppression of the sound speed.

The results shown in Figs.~\ref{fig:cs2} and~\ref{fig:cs2_k}
correspond to small initial field velocities, referred to as
soft initial conditions.
For hard initial conditions, characterized by larger initial field velocities,
the rest-frame sound speed exhibits significantly larger oscillation
amplitudes.
In this case, the pressure perturbations develop strong oscillations with
frequent zero crossings, whereas the corresponding density perturbations
remain comparatively small.
Consequently, the rest-frame sound speed,
\[
\hat{c}_s^2 \equiv
\frac{\delta \hat{P}_{\mathrm{DE}}}
{\delta \hat{\rho}_{\mathrm{DE}}}\,,
\]
becomes highly oscillatory and can temporarily take negative values or exceed
unity.
Under these conditions, the heavy mode can no longer be neglected, and the
perturbation dynamics cannot be fully described by a simple effective
single-field description \cite{Shiu:2011qw}.
Moreover, these initial conditions also enhance the oscillation amplitude of
the modified sound speed \(c_\text{s}\), reflecting the non-negligible influence of
the heavy mode on the light-sector dynamics (see
Fig.~\ref{fig:pert-evol}(b), upper panel).

In this section, we have demonstrated the consistency between the numerical
solutions of the exact perturbation equations and the analytical description
provided by the modified equations.

In the next section, we analyze the evolution of the scalar-field
perturbations and investigate the impact of the model on the matter power
spectrum under different scenarios and initial conditions.

\section{Dark-energy clustering mechanisms}\label{clustering}
A direct analysis of scalar perturbations along the tangential and normal
directions provides valuable insight into the sub-horizon evolution of
dark-energy perturbations.

\begin{figure*}[!t]
    \centering
    \begin{minipage}[t]{0.325\textwidth}
        \includegraphics[width=\textwidth]{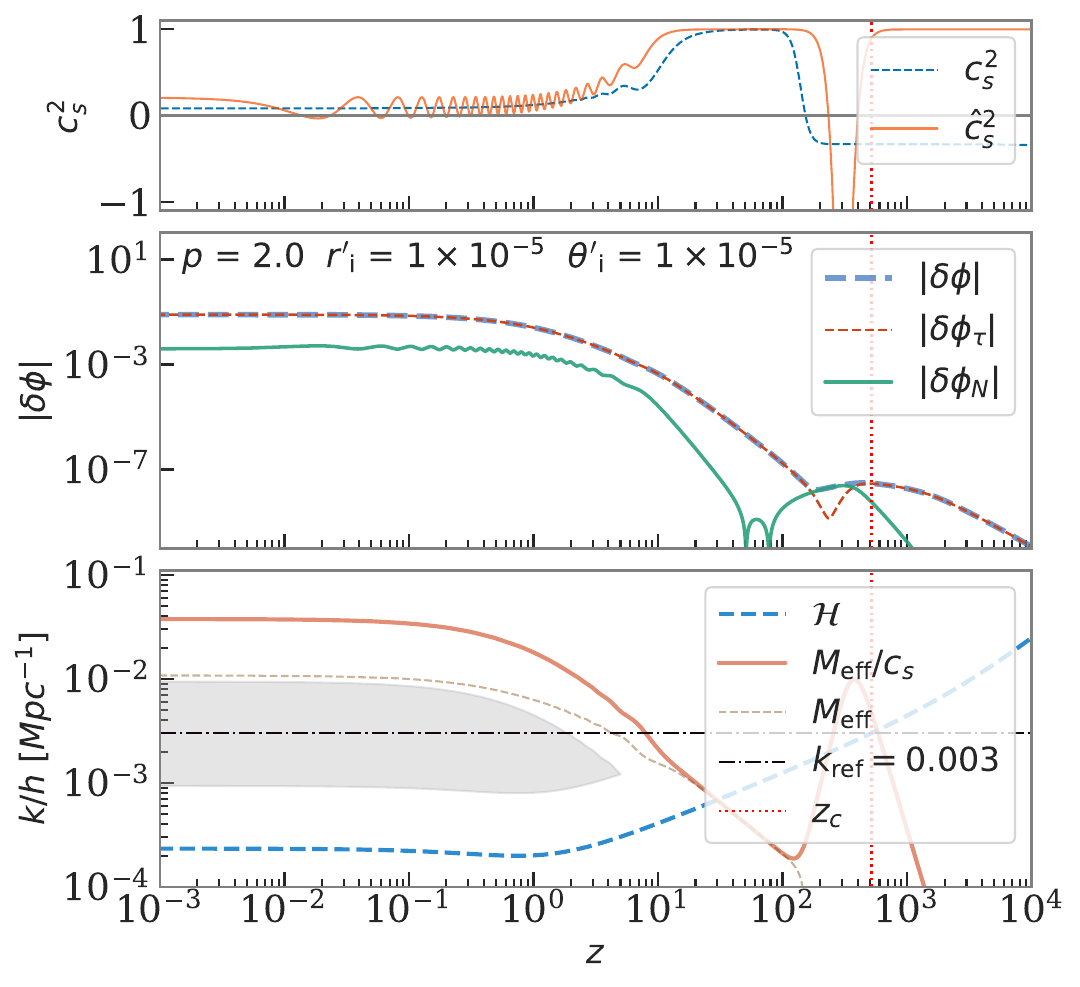}
        \centering
        {\small (a) Mechanism~1: Suppression of the modified sound speed of the light mode.}
        \label{fig:soft-pert}
    \end{minipage}
    \hfill
    \begin{minipage}[t]{0.325\textwidth}
        \includegraphics[width=\textwidth]{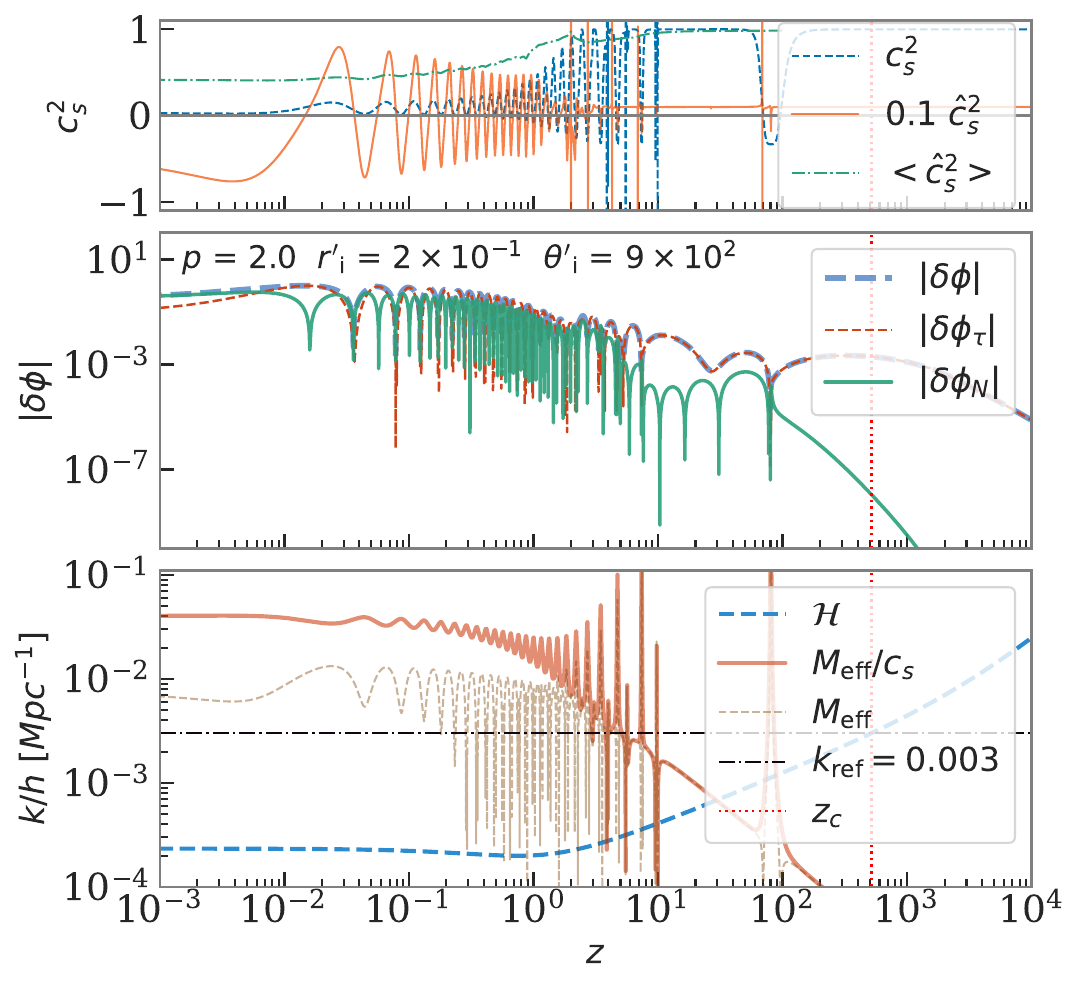}
        \centering
        {\small (b) Mechanism~2: Dynamically active heavy mode.}
        \label{fig:hard-pert}
    \end{minipage}
    \hfill
    \begin{minipage}[t]{0.325\textwidth}
        \includegraphics[width=\textwidth]{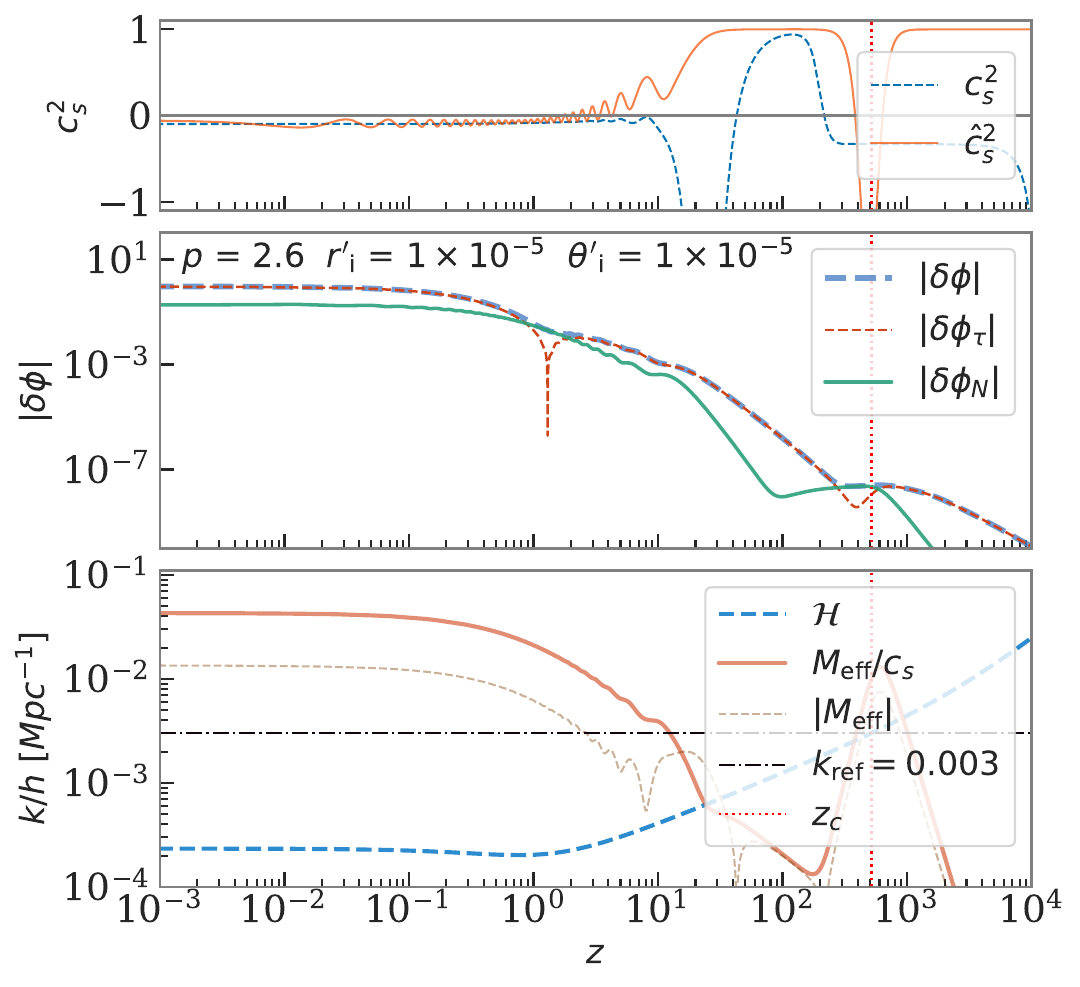}
        \centering
        {\small (c) Mechanism~3: Tachyonic instability from field-space curvature.}
        \label{fig:tachyon-pert}
    \end{minipage}

\caption{
Evolution of scalar-field perturbations for two sets of initial conditions and two values of the field-space curvature.
The parameters are \(\alpha = 0.002H_0^2\), \(m = 90H_0\), and \(r_0 = r_i = 7\times10^{-4}\).
In each subfigure, the upper panels show the sound speeds squared, including the modified sound speed squared \(c_\text{s}^2\), the rest-frame sound speed squared \(\hat{c}_s^2\), and, for panel (b), the averaged rest-frame sound speed squared \(\langle \hat{c}_s^2 \rangle\).
The middle panel displays the tangential and normal field perturbations \(\delta\phi^{\Ta}\) (dashed),
\(\delta\phi^{\N}\) (solid), and the total perturbation
\(|\delta\phi|=\sqrt{|\delta\phi^{\Ta}|^2+|\delta\phi^{\N}|^2}\) (thick dashed), all in reduced Planck units (\(M_{\rm{Pl}}=1\)).
The lower panel shows the background quantities \(\mathcal{M}_{\rm eff}\), \(\mathcal{H}\) (blue dashed) and
\(\mathcal{M}_{\rm eff}/c_\text{s}\) (orange solid), all in units of \(H_0\).
All results are shown for the reference comoving wavenumber
\(k_{\rm ref}=0.003\;h\,\mathrm{Mpc}^{-1}\) (black dash--dot line).
}
\label{fig:pert-evol}
\end{figure*}

\begin{figure*}[!t]
    \centering
    \begin{minipage}[t]{0.32\textwidth}
        \includegraphics[width=\textwidth]{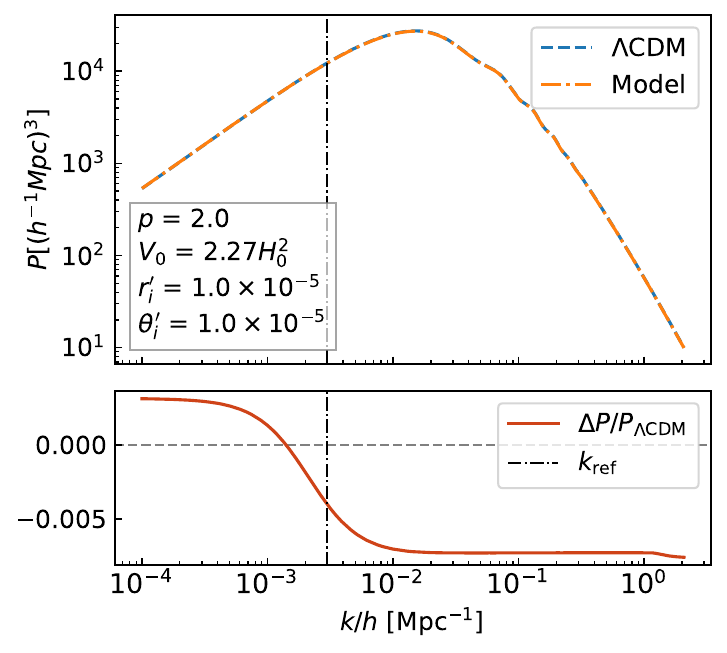}
        \centering
        {\small (a) Mechanism~1: Suppression of the modified sound speed of the light mode.}
        \label{fig:soft-power}
    \end{minipage}
    \hfill
    \begin{minipage}[t]{0.32\textwidth}
        \includegraphics[width=\textwidth]{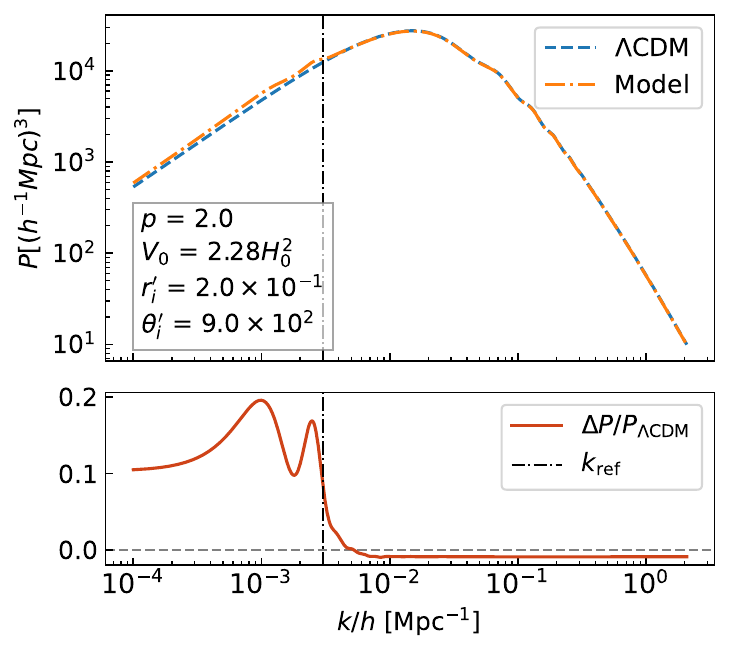}
        \centering
        {\small (b) Mechanism~2: Dynamically active heavy mode.}
        \label{fig:hard-power}
    \end{minipage}
    \hfill
    \begin{minipage}[t]{0.32\textwidth}
        \includegraphics[width=\textwidth]{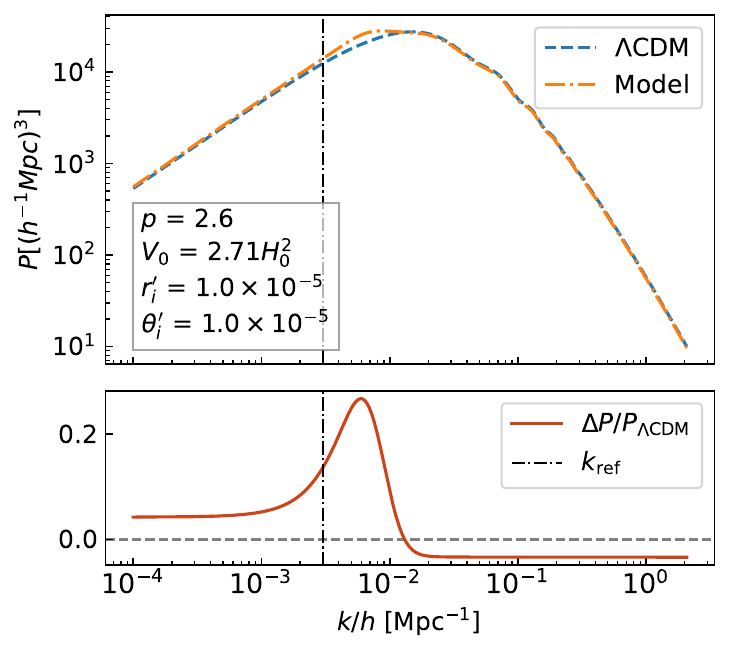}
        \centering
        {\small (c) Mechanism~3: Tachyonic instability from field-space curvature.}
        \label{fig:tachyon-power}
    \end{minipage}
    \caption{
Matter power spectra corresponding to the three mechanisms illustrated in Fig.~\ref{fig:pert-evol}. The lower panels show the relative difference with respect to \(\Lambda\mathrm{CDM}\).
}
    \label{fig:MPS}
\end{figure*}

In multifield models, several mechanisms can lead to dark-energy clustering or produce distinctive imprints on large-scale structure \cite{Akrami:2018ylq}.

\textit{1---Suppression of the modified sound speed of the light mode:}
In models with a non-geodesic (turning) background trajectory, the coupling between
tangential and normal fluctuations modifies the dispersion relation of the light mode,
leading to a reduced modified sound speed squared, \(c_\text{s}^2\ll1\).
In regimes where the heavy mode can be consistently integrated out,
dark-energy perturbations may, in principle, cluster on sub-horizon scales,
in contrast to standard single-field quintessence (Fig.~\ref{fig:soft-power}(a)).

\textit{2---Dynamically active heavy mode:}
Depending on the initial conditions, the heavy mode may remain dynamically relevant
at late times rather than being effectively suppressed.
In this case, the heavy mode behaves analogously to a canonical massive scalar field.
Once its effective mass exceeds the Hubble scale,
\(\mathcal{M}_{\rm eff}\gtrsim\mathcal{H}\),
the field undergoes oscillations and can contribute to clustering on sub-horizon scales (Fig.~\ref{fig:hard-power}(b)).

\textit{3---Tachyonic instability from field-space curvature:}
The curvature of the scalar-field manifold contributes to the effective mass of the heavy mode.
A sufficiently negative field-space curvature can render the heavy mode tachyonic.
In the late Universe, a mild instability of this kind may imprint distinctive features on structure formation,
potentially providing a probe of the underlying field-space geometry (Fig.~\ref{fig:tachyon-power}(c)).

In this section, we examine these three mechanisms and their impact on the matter power spectrum.

The different components of the perturbation evolution are shown in Fig.~\ref{fig:pert-evol}.
The upper panels display the evolution of the modified and rest-frame sound speeds.
The middle panels show the evolution of the scalar-field perturbations from
early times down to the present epoch (\(z\sim0\)), including both the
tangential component \(\delta\phi^{\Ta}\) and the normal component
\(\delta\phi^{\N}\), evaluated at the reference comoving wavenumber
\(k_{\rm ref}=0.003 \,h\,\mathrm{Mpc}^{-1}\).
The lower panels present the evolution of the background quantities,
including the effective mass and the Hubble parameter.
Together, these panels illustrate how the background evolution governs the corresponding perturbation dynamics.

The impact of these three mechanisms on structure formation is illustrated in
Fig.~\ref{fig:MPS}.
The upper panels compare the matter power spectrum of the model with that of
\(\Lambda\)CDM, while the lower panels show the fractional deviation from
\(\Lambda\)CDM,
\[
\frac{P_{\rm model}-P_{\Lambda{\rm CDM}}}{P_{\Lambda{\rm CDM}}}\,.
\]

\subsection{Suppression of the effective sound speed}

In Fig.~\ref{fig:pert-evol}(a), the smooth growth of the scalar-field
perturbations is clearly visible.
We adopt \(r'=\theta'=10^{-5}\) as soft initial conditions, which ensure a
smooth evolution toward the late-time attractor solution without significant
independent excitation of the heavy mode.
As shown in the middle panel, the normal component
\(\delta\phi^{\N}\) contributes non-negligibly to the scalar-field dynamics
relative to the tangential component under these conditions.
This indicates that the normal component continues to influence the low-energy
dynamics of the scalar-field perturbations through its coupling to the
tangential component, resulting in a reduced modified sound speed \(c_\text{s}\).

In this case,
\(\mathcal{M}_{\rm eff}^2=a^2(V_{\N\N}-\Omega^2)\) for \(p=2\), so an
increasing turning rate \(\Omega\) reduces
\(\mathcal{M}_{\rm eff}\) and enhances the coupling between tangential and
normal fluctuations.
Although the normal component remains coupled and influences the low-energy
dynamics through the modified sound speed, the corresponding heavy mode does
not propagate independently.

More precisely, based on the background and perturbation evolution shown in
Fig.~\ref{fig:pert-evol}(a), once the evolution enters the sub-horizon regime
(\(k>\mathcal{H}\)) and for redshifts \(z<z_c\), the effective mass
\(\mathcal{M}_{\rm eff}\) exceeds the Hubble parameter
\(\mathcal{H}\) and continues to grow, while the modified sound speed
\(c_\text{s}^2\) decreases.
After the condition
\(\mathcal{M}_{\rm eff}/c_\text{s}>k_{\rm ref}=0.003\,h\,\mathrm{Mpc}^{-1}\) is satisfied, the system
remains in a regime where the heavy-mode frequency satisfies
\(\omega_+^2\simeq\mathcal{M}_{\rm eff}^2/c_\text{s}^2\gg k_{\rm ref}^2\).
Consequently, as expected from the solutions of the modified perturbation
equations, the normal (heavy-field) component, together with the modified
sound speed, exhibits small-amplitude, high-frequency oscillations
characterized by
\(\mathcal{M}_{\rm eff}^2+4a^2\Omega^2
=\mathcal{M}_{\rm eff}^2/c_\text{s}^2\),
as shown in the middle panel of Fig.~\ref{fig:pert-evol}(a).
This behavior indicates that the heavy mode is not dynamically excited and
can therefore be consistently integrated out.

Even when the modified sound speed is strongly suppressed, the impact of
dark-energy clustering is limited by the finite cosmic time available before
the present epoch.
As a result, dark-energy perturbations have only a limited interval over which
they can grow within the range of scales where dark-energy clustering is
allowed, as defined by Eq.~(\ref{heavyMode_ineq}) for this mechanism.
Consequently, although a small sound speed permits clustering on
sub-horizon scales in principle, the limited time available before the
present epoch prevents this clustering from accumulating to a significant
level in the soft-initial-condition regime.

Therefore, for the chosen parameters and initial conditions, dark-energy
clustering driven solely by sound-speed suppression is not expected to produce
appreciable deviations at the present epoch, although it may become more
important at later cosmological times.
This conclusion is supported by Fig.~\ref{fig:MPS}(a), which shows that the
resulting matter power spectrum exhibits no significant deviation from the
\(\Lambda\)CDM prediction.%
\footnote{As mentioned earlier, the initial values of the scalar-field
perturbations, including both the tangential and normal components and their
velocities, are set to zero.
With this choice, the subsequent evolution of the scalar-field perturbations
is sourced by gravitational potential perturbations, which drive the system
away from the trivial solution through their coupling to the scalar-field
perturbation equations.}

It should be noted that exactly vanishing initial scalar-field velocities
cannot be imposed, since the tangential--normal decomposition becomes
ill-defined when the background field velocity vanishes
\cite{Gordon:2000hv}.
In practice, very small but nonzero initial velocities are adopted.
However, such choices can artificially enhance the initial value of the
turning rate \(\Omega\) (see Eq.~\ref{Omega_eq}) in the soft-initial-condition
regime.
This enhancement may leave imprints on the matter power spectrum.
Nevertheless, because it originates from the sensitivity to the initial
conditions rather than from an independent physical mechanism, we do not
consider it to constitute a separate clustering mechanism.

\subsection{Dynamically active heavy mode}\label{hard-IC}

The excitation of the heavy mode can play a key role in shaping the clustering
signatures of dark energy. A natural way to excite the heavy mode is through
hard initial conditions, implemented by assigning sufficiently large initial
velocities to the scalar field~\cite{Gao:2012uq}.

In the background evolution shown in Fig.~\ref{fig:w_vs_z}, hard initial
conditions cause the scalar field to oscillate around its background
trajectory, with an amplitude that depends on the magnitude of the initial
velocities \cite{Shiu:2011qw}. These oscillations are gradually damped by
Hubble friction, and the field eventually relaxes toward the late-time
attractor solution.

In the perturbation evolution, hard initial conditions lead to a direct
excitation of the heavy mode. As shown in Fig.~\ref{fig:pert-evol}(b), we adopt
significantly larger initial velocities for the scalar field,
\(r'_i = 10^{-1}\) and \(\theta'_i = 9\times10^{2}\), than in the
soft-initial-condition case. The resulting large initial kinetic energy
substantially enhances the contribution of the normal (heavy-field) component,
which manifests itself through an increased amplitude and pronounced
oscillatory behavior of the perturbations~\cite{Gao:2012uq}. This behavior
contrasts sharply with the soft-initial-condition mechanism shown in
Fig.~\ref{fig:pert-evol}(a).

As a consequence, the heavy mode becomes dynamically excited and can no longer
be integrated out, so the scalar-field system cannot be accurately described
within an effective single-field framework. This leads to enhanced,
strongly oscillatory features in the total scalar-field perturbation,
\[
|\delta\phi| = \sqrt{ |\delta\phi^{\Ta}|^{2} + |\delta\phi^{\N}|^{2} }\,.
\]

In this mechanism, the behavior is analogous to that of a massive scalar field
in cosmology: as the field undergoes coherent oscillations, clustering occurs
on scales exceeding the Jeans length \cite{Nambu:1989kh}.
In our case, although the background equation of state remains close to
\(w\simeq-1\) and the system remains in the dark-energy regime with
\(\mathcal M_{\rm eff}\equiv10^2\cH_0\) at the present epoch, the dominant
oscillation frequency is set by the heavy-mode frequency \(\omega_+\), with an
effective mass given by \(\mathcal{M}_{\mathrm{eff}}/c_\text{s}\).
As shown in Fig.~\ref{fig:pert-evol}(b), once this mass satisfies
\(\mathcal M_{\rm eff}/c_\text{s} \gtrsim \cH\), the heavy mode undergoes
oscillations, and its perturbations can contribute to gravitational clustering
for modes satisfying \(k < \mathcal M_{\mathrm{eff}}/c_\text{s}\), corresponding to
scales larger than the Compton wavelength.
These dynamics translate into a localized amplification of the matter power
spectrum on scales satisfying \(k < \mathcal M_{\mathrm{eff}}/c_\text{s}\), as shown
in Fig.~\ref{fig:MPS}(b) \cite{Amendola:2015ksp}.

Although under hard initial conditions the scalar-field system can no longer
be accurately described within an effective single-field theory, the
rest-frame sound speed remains suppressed on average. This behavior is
illustrated in Fig.~\ref{fig:pert-evol}(b), which shows the moving average of
the rest-frame sound speed,
\(\langle \hat{c}_s^2 \rangle\), for hard initial
conditions.\footnote{The plotted curve is obtained using a simple moving
average of the oscillatory rest-frame sound speed. The true level of
suppression may be stronger than suggested by the curve.}

It should be noted that, for these hard initial conditions, although the
linear calculation reliably identifies the range of scales over which
dark-energy clustering leaves an imprint on the matter power spectrum, the
precise amplitude of the deviation in this regime should be regarded as
indicative rather than exact, since
\(\delta\rho_{\rm DE}/\rho_{\rm DE}\) transiently approaches or even exceeds
unity near the present epoch. A fully nonlinear treatment may modify the
detailed shape of the deviation from \(\Lambda\)CDM without changing the
characteristic scale at which it occurs.

\subsection{Tachyonic instability from field-space curvature}
\label{tachyon-IC}

\begin{figure*}[!]
    \centering

    \begin{minipage}[t]{0.49\textwidth}
        \centering
        \includegraphics[width=\textwidth]{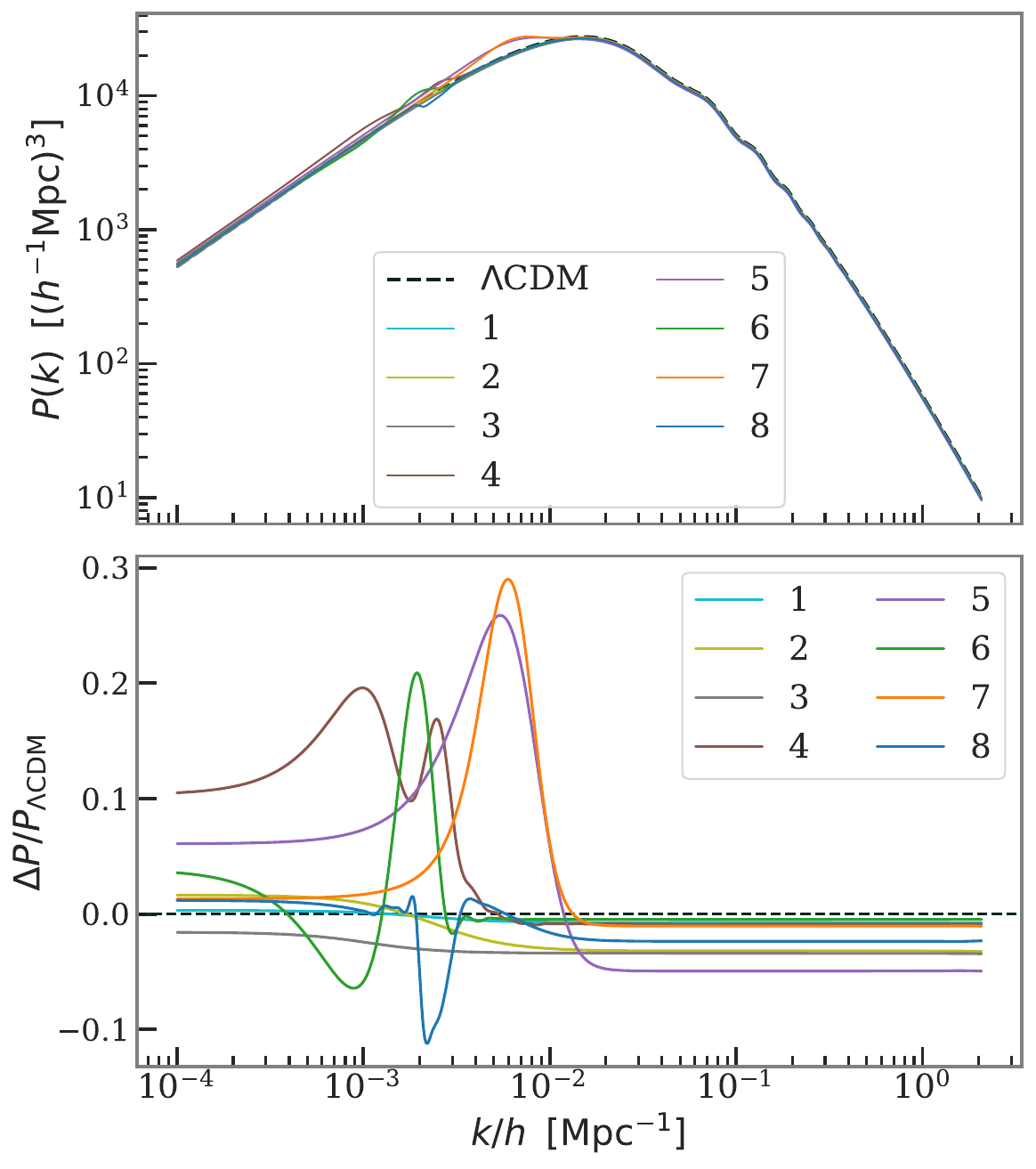}
        {\small (a) Matter power spectrum (models vs.\ $\Lambda$CDM).}
    \end{minipage}
    \hfill
    \begin{minipage}[t]{0.49\textwidth}
        \centering
        \includegraphics[width=\textwidth]{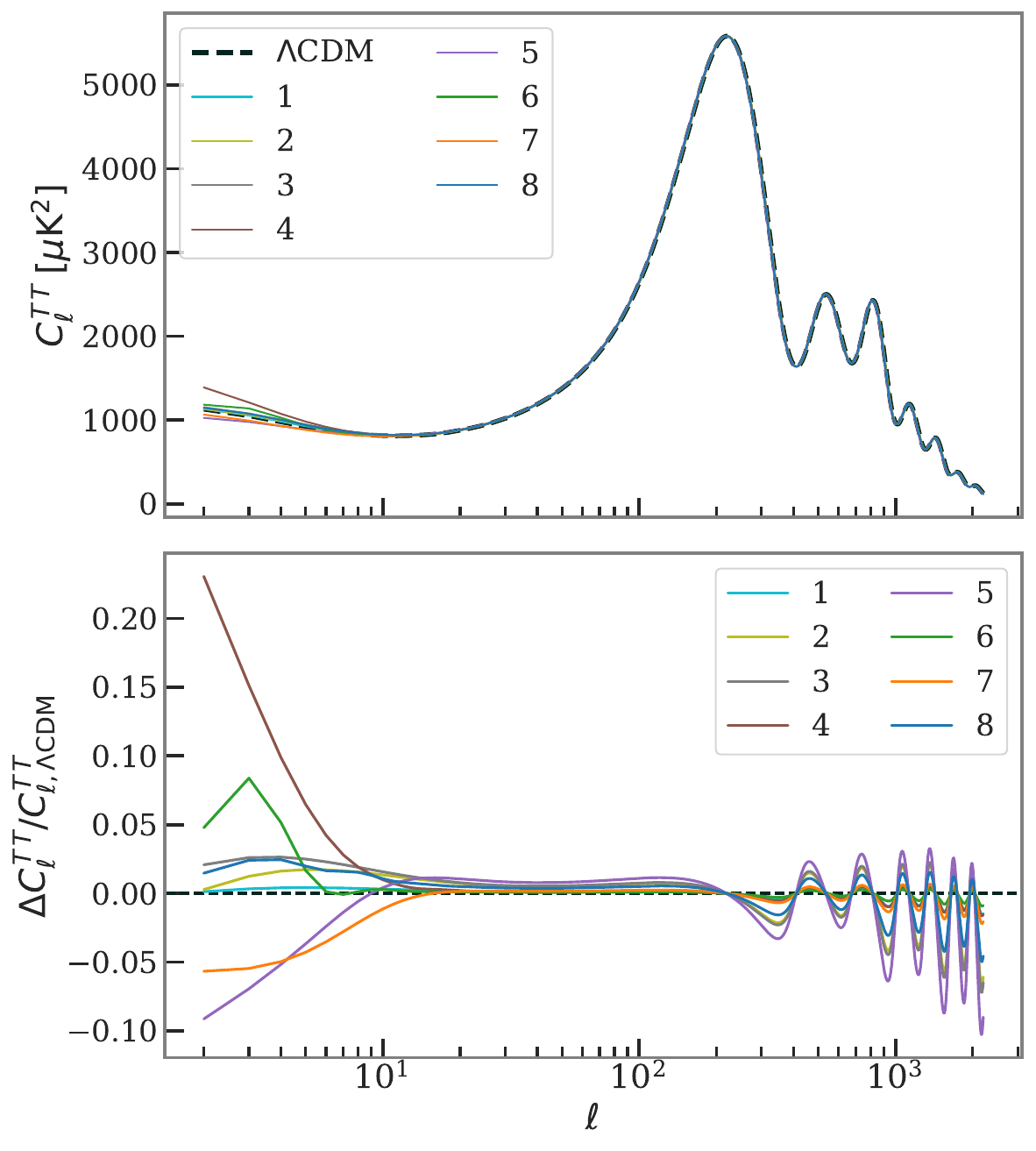}
        {\small (b) CMB temperature angular power spectrum (models vs.\ $\Lambda$CDM).}
    \end{minipage}

    \caption{Comparison with $\Lambda$CDM. (a) Matter power spectrum. (b) CMB temperature angular power spectrum.}
    \label{fig:mps-cltt}
\end{figure*}

\begin{table*}[!]
    \centering
    \includegraphics[width=\textwidth]{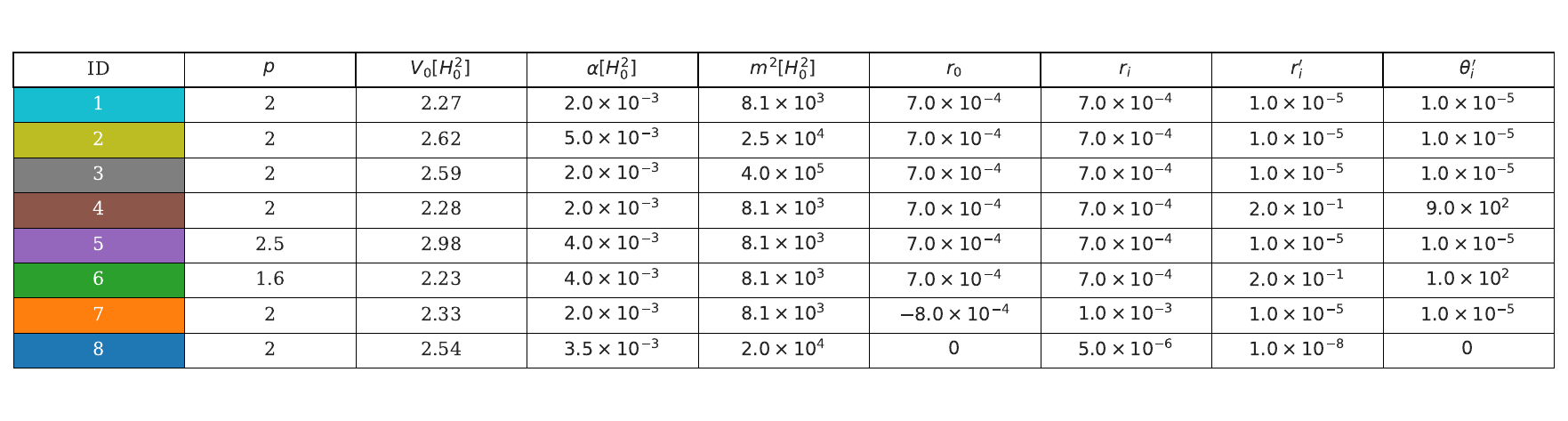}
    \caption{Model parameters used in Fig.~\ref{fig:mps-cltt}.}
    \label{tab:params}
\end{table*}

The third mechanism that can lead to dark-energy clustering or other distinctive
features in the matter power spectrum is a tachyonic instability sourced by a
sufficiently negative field-space curvature.
In our two-field dark-energy model, negative curvature (corresponding to
\(p>2\)) contributes to the effective mass squared of the normal component,
\(\mathcal{M}_{\rm eff}^2\), and can drive it to negative values during the
evolution.

In general, in the absence of additional effects, a negative
\(\mathcal{M}_{\rm eff}^2\) signals a tachyonic instability for perturbations
on scales where spatial gradients are subdominant, i.e., for wavenumbers
satisfying
\(k^2 \lesssim |\mathcal{M}_{\rm eff}^2|\).
However, in the presence of a sufficiently large turning rate \(\Omega\), the
dynamics are substantially modified.
In particular, under the conditions
\(k^2 < -\mathcal{M}_{\rm eff}^2 \ll 4a^2\Omega^2\),
the sound speed squared \(c_\text{s}^2\) becomes negative, and the tachyonic behavior
manifests itself in the effective description of the light mode.
Indeed, although \(\mathcal{M}_{\rm eff}^2<0\), the heavy-mode frequency
squared,
\(\omega_+^2=\mathcal{M}_{\rm eff}^2/c_\text{s}^2\),
remains positive.
As a result, the heavy mode continues to oscillate rather than undergoing
exponential growth.
By contrast, the light mode acquires a negative effective sound speed squared,
corresponding to an imaginary propagation speed and leading to the growth of
perturbations on sufficiently large scales.

These behaviors are illustrated in Fig.~\ref{fig:pert-evol}(c) for \(p=2.6\).
Here, we adopt soft initial conditions,
\(r'_i=\theta'_i=10^{-5}\), in order to isolate the effects of the tachyonic
instability from those associated with direct heavy-mode excitation.
Consistent with this picture, the impact on the matter power spectrum is most
pronounced for modes satisfying
\(k\lesssim|\mathcal{M}_{\rm eff}|\)
during the tachyonic phase, as shown in Fig.~\ref{fig:MPS}(c).

In the inflationary literature, this phenomenon is closely related to
\emph{geometrical destabilization}, in which an imaginary sound speed reflects
a transient tachyonic amplification rather than a fundamental pathology of the
theory~\cite{Renaux-Petel:2015mga, Garcia-Saenz_2018,
Garcia-Saenz:2018ifx}.
Although a tachyonic phase may be problematic in the late Universe if it
persists, a mild and transient instability can provide a valuable probe of the
underlying field-space curvature through distinctive features imprinted on
structure formation and, consequently, on the matter power spectrum.

Our computations indicate that, for the tachyonic mechanism to produce
noticeable deviations from the \(\Lambda\)CDM matter power spectrum,
dark-energy perturbations typically evolve into the nonlinear regime.
Consequently, the detailed evolution and precise amplitude of these
deviations may be affected by nonlinear effects, although the linear
calculation should still reliably identify the characteristic range of scales
over which the instability develops. Nevertheless, these results provide a
novel observational window into the geometry of field space in multifield
dark-energy models.

In this section, we have discussed the mechanisms that can produce deviations
in the matter power spectrum relative to the \(\Lambda\)CDM prediction and have
highlighted the distinctive features associated with each of them.
In the next section, we turn to the power spectra themselves and examine the
combined effects of the model parameters and initial conditions.
This allows us to categorize the resulting features of the power spectra
according to their underlying physical mechanisms.

\section{Observable consequences: power spectra}

In the previous section, we fixed the potential parameters in order to isolate
the effects of the initial conditions and the negative field-space curvature
on the matter power spectrum.
In this section, we investigate the combined effects of the model parameters,
the initial velocity conditions, and the field-space curvature on both the
matter power spectrum and the CMB temperature angular power spectrum.
We select the initial conditions and model parameters to cover all relevant
scenarios using a representative set of eight configurations, summarized in
Fig.~\ref{fig:mps-cltt} and Table~\ref{tab:params}.

\begin{figure}[!h]
    \noindent
    \includegraphics[width=\columnwidth]{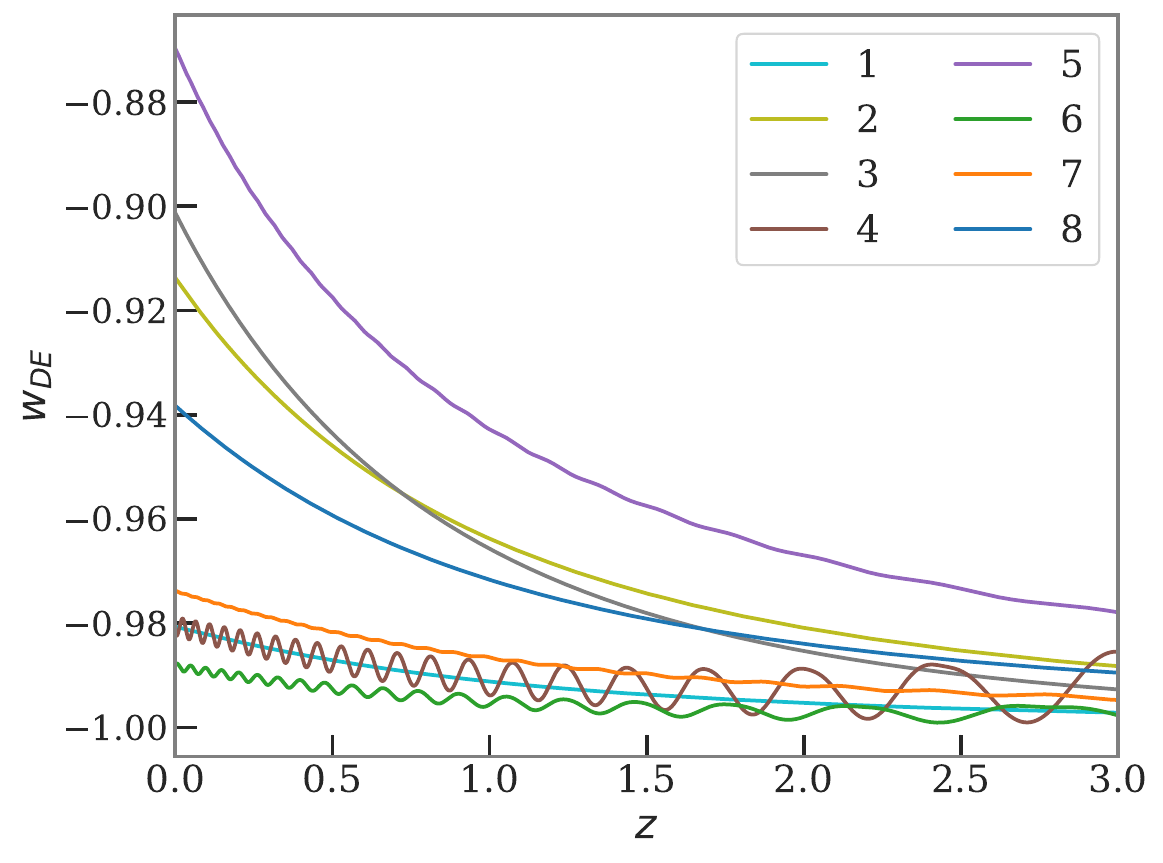}
    \caption{Evolution of the equation-of-state parameter \(w_{\rm DE}\) as a function of redshift \(z\) for the parameter sets listed in Table~\ref{tab:params}.}
    \label{EoSs}
\end{figure}

Curves~1 (discussed in the previous section), 2, and~3 correspond to the first
mechanism for dark-energy clustering, namely the suppression of the effective
sound speed. These curves exhibit the same overall behavior, with no
significant local deviation from the \(\Lambda\)CDM prediction for the chosen
parameters and initial conditions. Nevertheless, increasing the values of
\(\alpha\) and/or \(m\) leads to a noticeable shift, particularly at large
\(k\) and \(\ell\), while inducing only mild global changes on larger scales.
A more oscillatory behavior in the relative difference of the CMB temperature
power spectrum, together with a displacement of the matter power spectrum,
signals this shift for the first mechanism. This behavior can be attributed to
background effects projected onto the power spectra, arising from the increased
contribution of the scalar field to the background evolution for larger values
of the model parameters. Such a shift may be more consistent with the
observational data discussed in Section~\ref{sec:background}. This is evident
in Fig.~\ref{EoSs}, which shows the background evolution for each parameter set
and demonstrates that the shift in the power spectra is accompanied by a
deviation of the equation-of-state parameter toward values larger than
\(w=-1\).

Curve~4 corresponds to the second mechanism, in which hard initial conditions
excite the heavy mode, leading to localized enhancements in the matter power
spectrum, as discussed in detail in
Subsection~\ref{hard-IC}. It also exhibits a potentially observable imprint on
the CMB temperature power spectrum through a more pronounced enhancement at
small angular multipoles than in the other mechanisms.
In general, local deviations from the \(\Lambda\)CDM prediction induced by
clustering effects are more likely to be observed at low angular multipoles.

Curves~5 and~6 illustrate the effects of negative and positive field-space
curvature, respectively. Negative curvature (curve~5) leads to a tachyonic
instability of the heavy mode, as discussed in
Subsection~\ref{tachyon-IC}, resulting in localized enhancements of the matter
power spectrum. By contrast, for positive curvature, soft initial conditions
do not lead to significant deviations from the \(\Lambda\)CDM prediction.
However, clustering features emerge when hard initial conditions are employed,
even with relatively small initial velocities (curve~6). This indicates that,
for positive curvature, excitation of the heavy mode is necessary to produce
observable imprints on the matter power spectrum, and that positive curvature
enhances the clustering features induced by hard initial conditions.
Curve~6 can therefore be classified as an example of the second mechanism.

Curve~7 illustrates the role of an unconventional choice of \(r_0\), for which it is
assigned a negative value despite the use of a power-law field-space metric.
Although negative values of \(r\) are forbidden, the specific dynamics of the
model ensure that the field evolves toward the equilibrium value
\(r_{\rm eq}\) at late times, rather than toward \(r_0\) as a vacuum
expectation value. Consequently, \(r\) never becomes negative, and the
numerical computation remains well behaved. This setup can potentially produce
observable imprints on the matter power spectrum, with a pattern similar to
that of the (potentially transient) tachyonic-instability mechanism, since
\(\mathcal{M}_{\rm eff}^2\) becomes negative.

Finally, curve~8 illustrates the effects of an unphysical scenario in which
the initial values of the scalar fields are chosen to be close to zero. As
discussed in the previous section, this choice is not physically motivated.

\section{Conclusions}
\label{sec:conclusions}

In this work, we have investigated the cosmological implications of a two-field
dark-energy model with a curved field-space metric, focusing on both the
background evolution and the linear perturbation dynamics. By implementing the
full set of background and perturbation equations in a modified version of
\texttt{CAMB}, we were able to solve the system numerically and compute the
resulting matter power spectrum and CMB temperature angular power spectrum
within a self-consistent framework.

At the background level, we found that the model admits viable late-time
accelerating solutions over a broad range of parameters. The non-geodesic (turning) field-space trajectories allow the dark-energy
equation-of-state parameter to remain close to \(w\simeq-1\) while still
accommodating deviations favored by recent observational indications. In particular,
variations of the potential parameters \(m\) and \(\alpha\) lead to controlled
departures from \(\Lambda\)CDM that are compatible with current background
constraints, supporting the phenomenological viability of multifield
dark-energy scenarios.

At the level of linear perturbations, we analyzed the dynamics using a
decomposition into tangential and normal components and compared the exact
numerical solutions with the effective single-field description derived in the
sub-horizon limit. We demonstrated excellent consistency between the modified
sound speed obtained from the effective theory and the gauge-invariant
rest-frame sound speed extracted from the full perturbation equations,
particularly for soft initial conditions. This agreement confirms the validity
of the effective single-field interpretation in regimes where the heavy mode
can be consistently integrated out.

We identified and systematically studied three distinct mechanisms that can
lead to dark-energy clustering or to observable imprints in the matter power
spectrum: suppression of the effective sound speed of the light mode,
dynamical excitation of the heavy mode due to hard initial conditions, and
tachyonic instability induced by negative field-space curvature. While
sound-speed suppression alone does not produce observable local deviations
from \(\Lambda\)CDM within the finite age of the Universe, excitation of the
heavy mode and tachyonic growth can generate localized enhancements in the
matter power spectrum together with corresponding features in the CMB
temperature anisotropies.

Overall, our results demonstrate that multifield dark-energy models with curved
field-space geometry can give rise to rich and testable phenomenology beyond
that of single-field quintessence while remaining consistent with current
observations. The framework developed here provides a robust numerical and
theoretical foundation for future studies, including parameter inference using
observational data and extensions to nonlinear structure formation.

\acknowledgments{}
Y.A.\ acknowledges support by the Spanish Research Agency (Agencia Estatal de Investigaci\'on)'s grant RYC2020-030193-I/AEI/10.13039/501100011033, by the European Social Fund (Fondo Social Europeo) through the Ram\'{o}n y Cajal programme within the State Plan for Scientific and Technical Research and Innovation (Plan Estatal de Investigaci\'on Cient\'ifica y T\'ecnica y de Innovaci\'on) 2017-2020, by the Spanish Research Agency through the grant IFT Centro de Excelencia Severo Ochoa No CEX2020-001007-S funded by MCIN/AEI/10.13039/501100011033, by the Spanish National Research Council (CSIC) through the Talent Attraction grant 20225AT025, and by the Spanish Research Agency's Consolidaci\'on Investigadora 2024 grant CNS2024-154430.

\newpage

\bibliography{bibliography.bib}

@article{SupernovaCosmologyProject:1998vns,
    author = "Perlmutter, S. and others",
    collaboration = "Supernova Cosmology Project",
    title = "{Measurements of $\Omega$ and $\Lambda$ from 42 high redshift supernovae}",
    eprint = "astro-ph/9812133",
    archivePrefix = "arXiv",
    primaryClass = "astro-ph",
    doi = "10.1086/307221",
    journal = "Astrophys. J.",
    volume = "517",
    pages = "565--586",
    year = "1999"
}

@article{SupernovaSearchTeam:1998fmf,
    author = "Riess, Adam G. and others",
    collaboration = "Supernova Search Team",
    title = "{Observational evidence from supernovae for an accelerating universe and a cosmological constant}",
    eprint = "astro-ph/9805201",
    archivePrefix = "arXiv",
    primaryClass = "astro-ph",
    doi = "10.1086/300499",
    journal = "Astron. J.",
    volume = "116",
    pages = "1009--1038",
    year = "1998"
}

@article{Planck:2018vyg,
    author = "Aghanim, N. and others",
    collaboration = "Planck",
    title = "{Planck 2018 results. VI. Cosmological parameters}",
    eprint = "1807.06209",
    archivePrefix = "arXiv",
    primaryClass = "astro-ph.CO",
    doi = "10.1051/0004-6361/201833910",
    journal = "Astron. Astrophys.",
    volume = "641",
    pages = "A6",
    year = "2020",
    note = "{[Erratum: Astron.Astrophys. 652, C4 (2021)]}"
}

@article{Martin:2012bt,
    author = "Martin, Jerome",
    title = "{Everything You Always Wanted To Know About The Cosmological Constant Problem (But Were Afraid To Ask)}",
    eprint = "1205.3365",
    archivePrefix = "arXiv",
    primaryClass = "astro-ph.CO",
    doi = "10.1016/j.crhy.2012.04.008",
    journal = "Comptes Rendus Physique",
    volume = "13",
    pages = "566--665",
    year = "2012"
}

@article{Burgess:2013ara,
    author = "Burgess, C. P.",
    title = "{The Cosmological Constant Problem: Why it's hard to get Dark Energy from Micro-physics}",
    eprint = "1309.4133",
    archivePrefix = "arXiv",
    primaryClass = "hep-th",
    reportNumber = "PI-COSMO-297",
    doi = "10.1093/acprof:oso/9780198728856.003.0004",
    journal = "Ox. U. Press",
    pages = "149--197",
    year = "2015"
}

@article{Copeland:2006wr,
    author = "Copeland, Edmund J. and Sami, M. and Tsujikawa, Shinji",
    title = "{Dynamics of dark energy}",
    eprint = "hep-th/0603057",
    archivePrefix = "arXiv",
    primaryClass = "hep-th",
    doi = "10.1142/S021827180600942X",
    journal = "Int. J. Mod. Phys. D",
    volume = "15",
    pages = "1753--1936",
    year = "2006"
}

@article{Akrami:2020zfz,
    author = "Akrami, Yashar and Sasaki, Misao and Solomon, Adam R. and Vardanyan, Valeri",
    title = "{Multi-field dark energy: cosmic acceleration on a steep potential}",
    eprint = "2008.13660",
    archivePrefix = "arXiv",
    primaryClass = "astro-ph.CO",
    reportNumber = "YITP-20-111",
    doi = "10.1016/j.physletb.2021.136427",
    journal = "Phys. Lett. B",
    volume = "819",
    pages = "136427",
    year = "2021"
}

@article{Eskilt:2022hug,
    author = "Eskilt, Johannes R. and Akrami, Yashar and Solomon, Adam R. and Vardanyan, Valeri",
    title = "{Cosmological dynamics of multifield dark energy}",
    eprint = "2201.08841",
    archivePrefix = "arXiv",
    primaryClass = "astro-ph.CO",
    doi = "10.1103/PhysRevD.106.023512",
    journal = "Phys. Rev. D",
    volume = "106",
    pages = "023512",
    year = "2022"
}

@ARTICLE{1995ApJ...455....7M,
       author = {{Ma}, Chung-Pei and {Bertschinger}, Edmund},
        title = "{Cosmological Perturbation Theory in the Synchronous and Conformal Newtonian Gauges}",
      journal = {\apj},
         year = 1995,
        month = dec,
       volume = {455},
        pages = {7},
          doi = {10.1086/176550},
archivePrefix = {arXiv},
       eprint = {astro-ph/9506072},
 primaryClass = {astro-ph},
       adsurl = {https://ui.adsabs.harvard.edu/abs/1995ApJ...455....7M}
}

@article{PhysRevD.69.083503,
  title = {Probing dark energy perturbations: The dark energy equation of state and speed of sound as measured by WMAP},
  author = {Bean, Rachel and Dor\'e, Olivier},
  journal = {Phys. Rev. D},
  volume = {69},
  issue = {8},
  pages = {083503},
  numpages = {8},
  year = {2004},
  month = {Apr},
  publisher = {American Physical Society},
  doi = {10.1103/PhysRevD.69.083503},
  url = {https://link.aps.org/doi/10.1103/PhysRevD.69.083503}
}

@article{Xian_Gao_2012,
doi = {10.1088/1475-7516/2012/10/040},
url = {https://doi.org/10.1088/1475-7516/2012/10/040},
year = {2012},
month = {oct},
publisher = {},
volume = {2012},
number = {10},
pages = {040},
author = {Xian Gao and David Langlois and Shuntaro Mizuno},
title = {Influence of heavy modes on perturbations in multiple field inflation},
journal = {Journal of Cosmology and Astroparticle Physics}
}

@article{PhysRevD.86.121301,
  title = {Heavy fields, reduced speeds of sound, and decoupling during inflation},
  author = {Ach\'ucarro, Ana and Atal, Vicente and C\'espedes, Sebasti\'an and Gong, Jinn-Ouk and Palma, Gonzalo A. and Patil, Subodh P.},
  journal = {Phys. Rev. D},
  volume = {86},
  issue = {12},
  pages = {121301},
  numpages = {5},
  year = {2012},
  month = {Dec},
  publisher = {American Physical Society},
  doi = {10.1103/PhysRevD.86.121301},
  url = {https://link.aps.org/doi/10.1103/PhysRevD.86.121301}
}

@article{Garcia-Saenz_2018,
doi = {10.1088/1475-7516/2018/11/005},
url = {https://doi.org/10.1088/1475-7516/2018/11/005},
year = {2018},
month = {nov},
publisher = {},
volume = {2018},
number = {11},
pages = {005},
author = {Garcia-Saenz, Sebastian and Renaux-Petel, S\'{e}bastien},
title = {Flattened non-Gaussianities from the effective field theory of inflation with imaginary speed of sound},
journal = {Journal of Cosmology and Astroparticle Physics}
}

@article{Shiu:2011qw,
    author = "Shiu, Gary and Xu, Jiajun",
    title = "{Effective Field Theory and Decoupling in Multi-field Inflation: An Illustrative Case Study}",
    eprint = "1108.0981",
    archivePrefix = "arXiv",
    primaryClass = "hep-th",
    doi = "10.1103/PhysRevD.84.103509",
    journal = "Phys. Rev. D",
    volume = "84",
    pages = "103509",
    year = "2011"
}

@article{Gordon:2000hv,
    author = "Gordon, Christopher and Wands, David and Bassett, Bruce A. and Maartens, Roy",
    title = "{Adiabatic and entropy perturbations from inflation}",
    eprint = "astro-ph/0009131",
    archivePrefix = "arXiv",
    doi = "10.1103/PhysRevD.63.023506",
    journal = "Phys. Rev. D",
    volume = "63",
    pages = "023506",
    year = "2000"
}

@article{Gao:2012uq,
    author = "Gao, Xian and Langlois, David and Mizuno, Shuntaro",
    title = "{Influence of heavy modes on perturbations in multiple field inflation}",
    eprint = "1205.5275",
    archivePrefix = "arXiv",
    primaryClass = "hep-th",
    doi = "10.1088/1475-7516/2012/10/040",
    journal = "JCAP",
    volume = "10",
    pages = "040",
    year = "2012"
}

@article{Lewis:1999bs,
    author = "Lewis, Antony and Challinor, Anthony and Lasenby, Anthony",
    title = "{Efficient computation of CMB anisotropies in closed FRW models}",
    eprint = "astro-ph/9911177",
    archivePrefix = "arXiv",
    doi = "10.1086/309179",
    journal = "Astrophys. J.",
    volume = "538",
    pages = "473--476",
    year = "2000"
}

@article{Bedroya:2025fwh,
    author = "Bedroya, Alek and Obied, Georges and Vafa, Cumrun and Wu, David H.",
    title = "{Evolving Dark Sector and the Dark Dimension Scenario}",
    eprint = "2507.03090",
    archivePrefix = "arXiv",
    primaryClass = "astro-ph.CO",
    month = "7",
    year = "2025"
}

@article{Nambu:1989kh,
    author = "Nambu, Yasusada and Sasaki, Misao",
    title = "{Quantum Treatment of Cosmological Axion Perturbations}",
    reportNumber = "RRK-89-32",
    doi = "10.1103/PhysRevD.42.3918",
    journal = "Phys. Rev. D",
    volume = "42",
    pages = "3918--3924",
    year = "1990"
}

@article{Obied:2018sgi,
      author         = "Obied, Georges and Ooguri, Hirosi and Spodyneiko, Lev and
                        Vafa, Cumrun",
      title          = "{De Sitter Space and the Swampland}",
      year           = "2018",
      eprint         = "1806.08362",
      archivePrefix  = "arXiv",
      primaryClass   = "hep-th",
      reportNumber   = "CALT-TH-2018-020, IPMU18-0100",
      SLACcitation   = "%%CITATION = ARXIV:1806.08362;%%"
}

@article{Agrawal:2018own,
      author         = "Agrawal, Prateek and Obied, Georges and Steinhardt, Paul
                        J. and Vafa, Cumrun",
      title          = "{On the Cosmological Implications of the String
                        Swampland}",
      journal        = "Phys. Lett.",
      volume         = "B784",
      year           = "2018",
      pages          = "271-276",
      doi            = "10.1016/j.physletb.2018.07.040",
      eprint         = "1806.09718",
      archivePrefix  = "arXiv",
      primaryClass   = "hep-th",
      SLACcitation   = "%%CITATION = ARXIV:1806.09718;%%"
}

@article{Ooguri:2018wrx,
      author         = "Ooguri, Hirosi and Palti, Eran and Shiu, Gary and Vafa,
                        Cumrun",
      title          = "{Distance and de Sitter Conjectures on the Swampland}",
      journal        = "Phys. Lett.",
      volume         = "B788",
      year           = "2019",
      pages          = "180-184",
      doi            = "10.1016/j.physletb.2018.11.018",
      eprint         = "1810.05506",
      archivePrefix  = "arXiv",
      primaryClass   = "hep-th",
      SLACcitation   = "%%CITATION = ARXIV:1810.05506;%%"
}

@article{Akrami:2018ylq,
      author         = "Akrami, Yashar and Kallosh, Renata and Linde, Andrei and
                        Vardanyan, Valeri",
      title          = "{The Landscape, the Swampland and the Era of Precision
                        Cosmology}",
      journal        = "Fortsch. Phys.",
      volume         = "67",
      year           = "2019",
      number         = "1-2",
      pages          = "1800075",
      doi            = "10.1002/prop.201800075",
      eprint         = "1808.09440",
      archivePrefix  = "arXiv",
      primaryClass   = "hep-th",
      SLACcitation   = "%%CITATION = ARXIV:1808.09440;%%"
}

@article{Raveri:2018ddi,
      author         = "Raveri, Marco and Hu, Wayne and Sethi, Savdeep",
      title          = "{Swampland Conjectures and Late-Time Cosmology}",
      journal        = "Phys. Rev.",
      volume         = "D99",
      year           = "2019",
      number         = "8",
      pages          = "083518",
      doi            = "10.1103/PhysRevD.99.083518",
      eprint         = "1812.10448",
      archivePrefix  = "arXiv",
      primaryClass   = "hep-th",
      reportNumber   = "EFI-18-22",
      SLACcitation   = "%%CITATION = ARXIV:1812.10448;%%"
}

@article{Garcia-Saenz:2018ifx,
    author = "Garcia-Saenz, Sebastian and Renaux-Petel, S\'ebastien and Ronayne, John",
    title = "{Primordial fluctuations and non-Gaussianities in sidetracked inflation}",
    eprint = "1804.11279",
    archivePrefix = "arXiv",
    primaryClass = "astro-ph.CO",
    doi = "10.1088/1475-7516/2018/07/057",
    journal = "JCAP",
    volume = "07",
    pages = "057",
    year = "2018"
}

@article{Achucarro:2018vey,
    author = "Ach{\'u}carro, Ana and Palma, Gonzalo A.",
    title = "{The string swampland constraints require multi-field inflation}",
    eprint = "1807.04390",
    archivePrefix = "arXiv",
    primaryClass = "hep-th",
    doi = "10.1088/1475-7516/2019/02/041",
    journal = "JCAP",
    volume = "02",
    pages = "041",
    year = "2019"
}

@article{Achucarro:2012sm,
    author = "Achucarro, Ana and Gong, Jinn-Ouk and Hardeman, Sjoerd and Palma, Gonzalo A. and Patil, Subodh P.",
    title = "{Effective theories of single field inflation when heavy fields matter}",
    eprint = "1201.6342",
    archivePrefix = "arXiv",
    primaryClass = "hep-th",
    reportNumber = "CERN-PH-TH-2011-222, CPHT-RR-055.0711, LPTENS-11-27",
    doi = "10.1007/JHEP05(2012)066",
    journal = "JHEP",
    volume = "05",
    pages = "066",
    year = "2012"
}

@article{Achucarro:2010jv,
    author = "Achucarro, Ana and Gong, Jinn-Ouk and Hardeman, Sjoerd and Palma, Gonzalo A. and Patil, Subodh P.",
    title = "{Mass hierarchies and non-decoupling in multi-scalar field dynamics}",
    eprint = "1005.3848",
    archivePrefix = "arXiv",
    primaryClass = "hep-th",
    reportNumber = "CPHT-RR-039.0510, LPTENS-10-20",
    doi = "10.1103/PhysRevD.84.043502",
    journal = "Phys. Rev. D",
    volume = "84",
    pages = "043502",
    year = "2011"
}

@book{Amendola:2015ksp,
    author = "Amendola, Luca and Tsujikawa, Shinji",
    title = "{Dark Energy}: {Theory and Observations}",
    isbn = "978-1-107-45398-2",
    publisher = "Cambridge University Press",
    month = "1",
    year = "2015"
}

@article{Renaux-Petel:2015mga,
    author = "Renaux-Petel, S\'ebastien and Turzy\'nski, Krzysztof",
    title = "{Geometrical Destabilization of Inflation}",
    eprint = "1510.01281",
    archivePrefix = "arXiv",
    primaryClass = "astro-ph.CO",
    doi = "10.1103/PhysRevLett.117.141301",
    journal = "Phys. Rev. Lett.",
    volume = "117",
    number = "14",
    pages = "141301",
    year = "2016"
}

@article{Douglas:2006es,
    author = "Douglas, Michael R. and Kachru, Shamit",
    title = "{Flux compactification}",
    eprint = "hep-th/0610102",
    archivePrefix = "arXiv",
    reportNumber = "SLAC-PUB-12131",
    doi = "10.1103/RevModPhys.79.733",
    journal = "Rev. Mod. Phys.",
    volume = "79",
    pages = "733--796",
    year = "2007"
}

@article{Arvanitaki:2009fg,
    author = "Arvanitaki, Asimina and Dimopoulos, Savas and Dubovsky, Sergei and Kaloper, Nemanja and March-Russell, John",
    title = "{String Axiverse}",
    eprint = "0905.4720",
    archivePrefix = "arXiv",
    primaryClass = "hep-th",
    doi = "10.1103/PhysRevD.81.123530",
    journal = "Phys. Rev. D",
    volume = "81",
    pages = "123530",
    year = "2010"
}

@article{Brown:2017osf,
    author = "Brown, Adam R.",
    title = "{Hyperbolic Inflation}",
    eprint = "1705.03023",
    archivePrefix = "arXiv",
    primaryClass = "hep-th",
    doi = "10.1103/PhysRevLett.121.251601",
    journal = "Phys. Rev. Lett.",
    volume = "121",
    number = "25",
    pages = "251601",
    year = "2018"
}

@article{Garg:2018reu,
    author = "Garg, Sumit K. and Krishnan, Chethan",
    title = "{Bounds on Slow Roll and the de Sitter Swampland}",
    eprint = "1807.05193",
    archivePrefix = "arXiv",
    primaryClass = "hep-th",
    doi = "10.1007/JHEP11(2019)075",
    journal = "JHEP",
    volume = "11",
    pages = "075",
    year = "2019"
}

@article{Alestas:2024gxe,
    author = "Alestas, George and Delgado, Matilda and Ruiz, Ignacio and Akrami, Yashar and Montero, Miguel and Nesseris, Savvas",
    title = "{Is curvature-assisted quintessence observationally viable?}",
    eprint = "2406.09212",
    archivePrefix = "arXiv",
    primaryClass = "hep-th",
    reportNumber = "IFT-UAM/CSIC-24-83",
    doi = "10.1103/PhysRevD.110.106010",
    journal = "Phys. Rev. D",
    volume = "110",
    number = "10",
    pages = "106010",
    year = "2024"
}

@article{Akrami:2025zlb,
    author = "Akrami, Yashar and Alestas, George and Nesseris, Savvas",
    title = "{Has DESI detected exponential quintessence?}",
    eprint = "2504.04226",
    archivePrefix = "arXiv",
    primaryClass = "astro-ph.CO",
    reportNumber = "IFT-UAM/CSIC-25-36",
    month = "4",
    year = "2025"
}
\end{document}